\documentclass[%
aps,nofootinbib,notitlepage,superscriptaddress,10Limberkipt,prd, twocolumn,
 amsmath,amssymb
]{revtex4-2}
\usepackage[caption=false]{subfig}
\usepackage{braket}
\usepackage{float}
\usepackage{lipsum}
\usepackage{graphicx}
\usepackage{dcolumn}
\usepackage{bm}
\usepackage{natbib}
\usepackage{hyperref}
\hypersetup{
	colorlinks = true,
    linkcolor = Maroon,
    urlcolor  = Maroon,
    citecolor = Maroon
}
\usepackage{afterpage}
\usepackage{placeins}
\usepackage{microtype}
\usepackage{verbatim}
\usepackage[amssymb]{SIunits}
\usepackage{tabularx}
\usepackage[dvipsnames]{xcolor}
\usepackage{tikz}
\usepackage{mathtools}

\makeatletter
\let\@afterindenttrue\@afterindentfalse
\makeatother

\usepackage{color}
\definecolor{darkgreen}{RGB}{0,120,0}

\newcommand{\av}[1]{\langle{#1}\rangle}

\newcommand{\vP}{{\bm P}}
\newcommand{\be}{\begin{equation}}
\newcommand{\ee}{\end{equation}}
\newcommand{\bea}{\begin{eqnarray}}
\newcommand{\eea}{\end{eqnarray}}
\newcommand{\beas}{\begin{eqnarray*}}
\newcommand{\eeas}{\end{eqnarray*}}

\newcommand{\vx}{\mathbf{x}}
\newcommand{\vk}{\mathbf{k}}
\newcommand{\vq}{\mathbf{q}}
\newcommand{\vp}{\mathbf{p}}
\newcommand{\vu}{\mathbf{u}}
\newcommand{\vv}{\mathbf{v}}

\newcommand{\cov}{\mathrm{Cov}}

\newcommand{\bx}{{\boldsymbol x}}
\newcommand{\by}{{\boldsymbol y}}

\def\gsim{ \lower .75ex \hbox{$\sim$} \llap{\raise .27ex \hbox{$>$}} }
\def\lsim{ \lower .75ex \hbox{$\sim$} \llap{\raise .27ex \hbox{$<$}} }

\def\dalam{\hbox
{\vrule\vbox{\hrule\hbox to 1ex{ \hfill}\kern 1 ex\hrule}\vrule}}

\def\1/2{\hbox{$ {1 \over 2}$ }}

\newcommand{\columbia}{Department of Physics, Columbia University, New York, NY, USA 10027}
\newcommand{\uwmadison}{Department of Physics, University of Wisconsin-Madison, Madison, WI, USA 53706}
\newcommand{\caltech}{Department of Physics, California Institute of Technology, 1200 E. California Boulevard, Pasadena, CA 91125, USA}
\newcommand{\perimeter}{Perimeter Institute for Theoretical Physics, 31 Caroline St North, Waterloo, ON N2L 2Y5, Canada}

\begin{document}

\title{If at First You Don't Succeed, Trispectrum — I. Estimating the Matter Power Spectrum Covariance with Higher-Order Statistics}
\author{Samuel~Goldstein}
\email{sjg2215@columbia.edu}
\affiliation{\columbia}

\author{Kendrick M. Smith}
\affiliation{\perimeter} 

\author{Utkarsh Giri}
\affiliation{\caltech}

\author{Moritz M\"unchmeyer}
\affiliation{\uwmadison}

\begin{abstract}
\noindent 
We present a method to estimate non-Gaussian power spectrum covariance matrices by directly measuring the response of the small-scale power spectrum to long-wavelength perturbations via bispectrum and trispectrum estimators. Specifically, we derive estimators for the complete non-Gaussian matter power spectrum covariance, including the super-sample contribution, in terms of the squeezed bispectrum and collapsed trispectrum of the underlying density field. We apply these estimators to the \textsc{Quijote} simulations, and recover unbiased estimates of the small-scale ($k\gtrsim 0.15~h/{\rm Mpc}$) matter power spectrum covariance at the percent level using only 25 simulations -- comparable to the precision of the sample covariance estimated using 5,000 simulations. This technique significantly reduces the number of simulations needed to estimate power spectrum covariances and opens the possibility of inferring power spectrum covariances directly from survey data, enabling stringent tests of simulations and, potentially, power spectrum analyses that do not rely on external covariance matrices.
\end{abstract}

\maketitle

\section{Introduction}\label{Sec:Intro}

The power spectrum is the central summary statistic of modern cosmology. Extracting cosmological information from power spectrum measurements requires not only an unbiased estimate of the power spectrum itself, but also an accurate estimate of its covariance~\cite{Percival:2013sga}. On large scales, where density fluctuations are approximately Gaussian, the covariance takes a simple analytic form. On small scales, non-linear structure formation and astrophysical processes source non-Gaussianities in the density field, thus complicating the power spectrum covariance~\cite{Mohammed:2014lja}.

The non-Gaussian power spectrum covariance is set by the connected four-point function, \emph{i.e.}, the trispectrum. In the quasi-linear regime, where the trispectrum can be described perturbatively, analytic approaches yield fast and accurate estimates of the non-Gaussian covariance (see, \emph{e.g.}, Ref.~\cite{Bertolini:2015fya, Mohammed:2014lja, Wadekar:2019rdu, Nicola:2020lhi}); however, analytic methods inevitably break down on sufficiently small scales. To address these limitations, cosmological analyses typically rely on large ensembles of computationally expensive (and approximate) simulations to estimate the power spectrum covariance. Since each simulation provides only a single noisy realization of the power spectrum, and mode coupling can make the power spectrum covariance dense, modern surveys typically require thousands to tens of thousands of realizations for a stable covariance estimate. It would be useful to be able to estimate the power spectrum covariance using fewer simulations.

\begin{figure}[!t]
\centering
\includegraphics[width=0.99\linewidth]{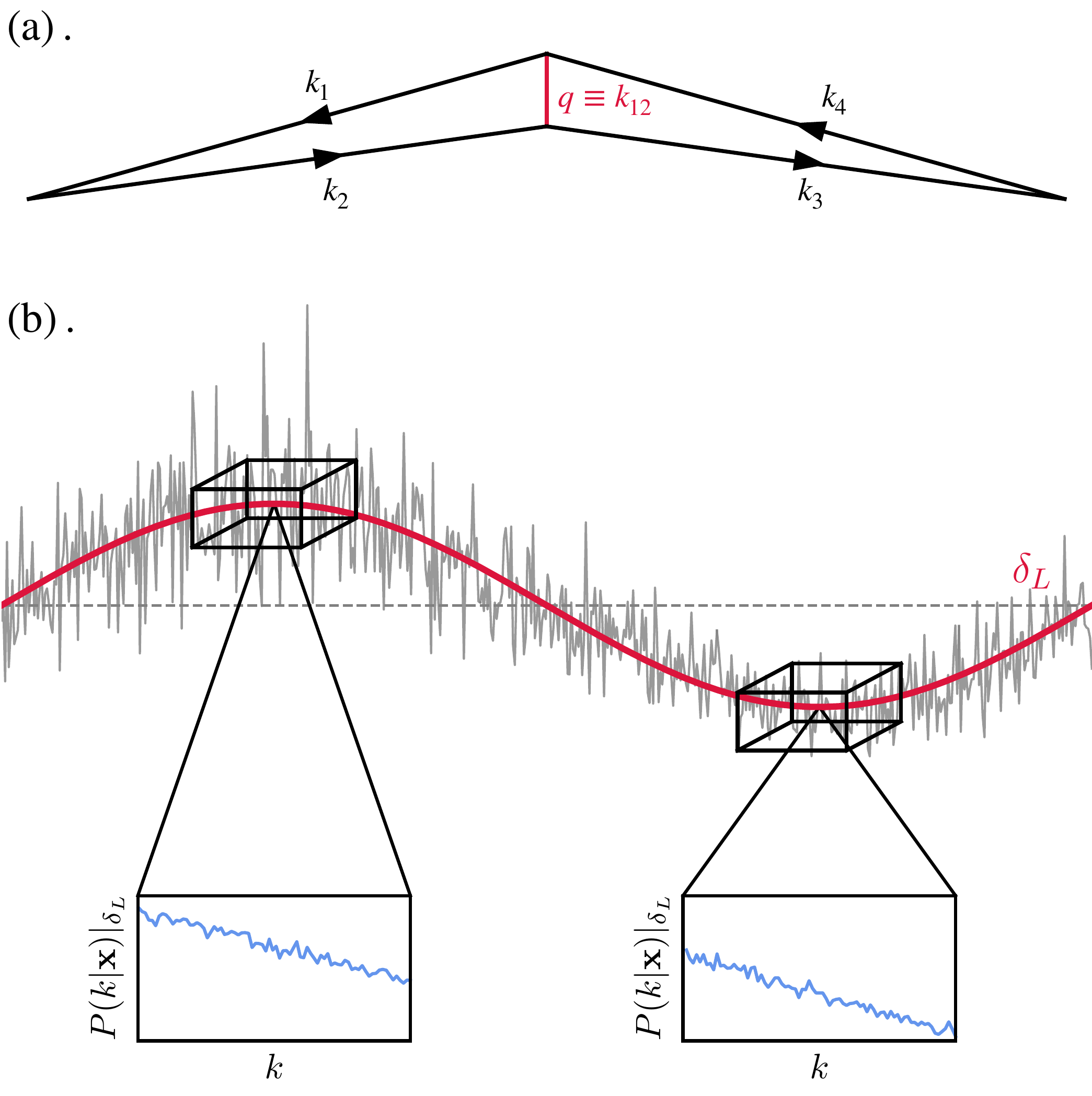}
\caption{{(a).} The kinematic configuration of the collapsed trispectrum in Fourier space. The collapsed limit corresponds to configurations where the internal momentum $q$ is much smaller than the external momenta $k_i$. (b). The configuration-space interpretation of the collapsed trispectrum. The collapsed trispectrum measures the variation in the locally measured small-scale power spectrum $P(k|\mathbf{x})|_{\delta_L}$ across spatial patches in the presence of a common long-wavelength mode $\delta_L$.} \label{fig:trispectrum_diagram}
\end{figure}

In this work, we present a new approach for covariance estimation based on directly measuring and fitting the power spectrum, squeezed bispectrum, and collapsed trispectrum of the field.\footnote{The squeezed bispectrum refers to the configuration of the three-point function $\langle \delta(\vk_1)\delta(\vk_2)\delta(\vk_3)\rangle'_c$ in which $k_1\ll k_2\approx k_3$. The collapsed trispectrum refers to the configuration of the connected four-point function  $\langle \delta(\vk_1)\delta(\vk_2)\delta(\vk_3)\delta(\vk_4)\rangle'_c$ in which $|\vk_1+\vk_2|\ll {\rm min}(k_1,k_2,k_3,k_4).$ } Intuitively, the squeezed bispectrum measures the response of the small-scale power spectrum to long-wavelength density fluctuations, and is therefore closely related to the super-sample covariance -- the covariance in the power spectrum due to the coupling of modes within the survey to modes larger than the survey volume. Similarly, the collapsed trispectrum measures the correlation between two locally measured small-scale power spectra that are coupled through a shared long mode (see Fig.~\ref{fig:trispectrum_diagram}). In the limit where the long mode goes to zero, the collapsed trispectrum yields the ``connected non-Gaussian covariance"\footnote{We follow the convention in the literature and split the non-Gaussian covariance into the connected non-Gaussian and super-sample terms, although we note that both terms are due to connected contributions to the four-point function.} -- the power spectrum covariance sourced purely by the mode coupling of small-scale density fluctuations. 

The utility of including bispectrum and trispectrum measurements to estimate the power spectrum covariance rests on two observations. First, for a given small-scale Fourier mode $k$ and a sufficiently large survey volume, there exist many squeezed bispectrum and collapsed trispectrum configurations with long-wavelength modes $q\ll k$. Each such long-wavelength mode provides an approximately independent probe of how the small-scale power spectrum responds to a long-wavelength density perturbation, allowing one to estimate the small-scale power spectrum covariance using many effective samples drawn from a single realization. Second, although the squeezed bispectrum and collapsed trispectrum can probe highly non-linear scales, they are strongly constrained by symmetries. Specifically, in the absence of primordial non-Gaussianity and equivalence-principle-violating physics, the long-wavelength dependence of the squeezed bispectrum and collapsed trispectrum is \emph{entirely} fixed by the so-called large-scale structure consistency relations~\cite{Kehagias:2013yd, Peloso:2013zw}. Indeed, Refs.~\cite{Goldstein:2022hgr, Giri:2023mpg, Goldstein:2023brb, Goldstein:2024bky, Goldstein:2025eyj, Zang:2025azh} used these relations to construct estimators for primordial non-Gaussianity in the non-linear regime. Here, we show that these non-perturbative primordial non-Gaussianity estimators can be repurposed for covariance estimation. 

Before proceeding, we discuss how our approach is related to existing power spectrum covariance estimation techniques. Conceptually, our approach is closely related to the response approach to the matter power spectrum covariance developed in Refs.~\cite{Barreira:2017sqa, Barreira:2017kxd, Barreira:2017fjz}, which expresses the covariance in terms of functions encoding the response of the power spectrum to long-wavelength perturbations. However, while Ref.~\cite{Barreira:2017kxd} estimates these response functions using separate universe simulations and perturbation theory, we compute these responses directly from the bispectrum and trispectrum of the survey. This obviates the need for specialized simulations and perturbative calculations, and, for sufficiently large survey volumes, could enable direct estimation of the non-Gaussian covariance from the data itself.

Our method is also closely related to ``internal covariance estimators," such as jackknife methods, which estimate the covariance directly from the data by partitioning the survey into sub-volumes~\cite{Escoffier:2016qnf, Euclid:2025fby}. Indeed, the collapsed trispectrum is effectively a jackknife power spectrum covariance estimator in disguise (see Fig.~\ref{fig:trispectrum_diagram}). However, unlike conventional jackknife approaches, which can be biased and often depend sensitively on the subdivision of the survey volume~\cite{Norberg:2008tg, Mohammad:2021aqc, Trusov:2023gsv}, the collapsed trispectrum naturally partitions the survey while consistently accounting for the physical correlations between long and short-wavelength modes. As a result, whereas existing internal covariance estimation methods can (erroneously) mix super-sample and connected non-Gaussian contributions to the covariance (see, \emph{e.g.}, Refs.~\cite{Lacasa:2017xbi, Trusov:2023gsv}), leading to biased covariance estimates, our approach makes the separation between (and estimation of) these contributions explicit. Furthermore, by jointly analyzing the long-wavelength power spectrum, squeezed bispectrum, and collapsed trispectrum, all of which trace the \emph{same} underlying density field, our approach yields precise estimates of the super-sample and connected non-Gaussian covariance that are not limited by the sample variance of the long-wavelength modes in the survey. Finally, our approach is formulated entirely in terms of Fourier-space $N$-point functions, providing a direct and physically transparent connection between the power spectrum covariance and mode-coupling in the underlying density field.

The remainder of this paper is organized as follows. In Sec.~\ref{Sec:background}, we present theoretical background on power spectrum covariance estimation and its connection to soft limits of the bispectrum and trispectrum. In Sec.~\ref{Sec:methodology}, we discuss the simulations, estimators, and likelihood used to reconstruct the non-Gaussian covariance. In Sec.~\ref{Sec:results}, we apply and validate our method using the \textsc{Quijote} simulations. We conclude in Sec.~\ref{Sec:Conclusions}. The appendices include additional details concerning the estimators and likelihood used in this work.

\section{Theoretical Background}\label{Sec:background}

In this section, we summarize the theoretical background required for our analysis. After establishing notation, we review the large-scale structure consistency relations that govern the behavior of the squeezed bispectrum and collapsed trispectrum. We then summarize our power spectrum, bispectrum, and trispectrum estimators. Finally, we derive the general structure of the three-dimensional (3D) power spectrum covariance,\footnote{In this work, we use ``real space" to refer to generic three-dimensional fields without redshift-space distortions. When distinguishing between position and momentum space, we use the terms configuration space and Fourier space.} emphasizing its connection to the squeezed bispectrum and collapsed trispectrum estimators. Although the power spectrum covariance has been studied extensively in the literature (see, \emph{e.g.}, Ref.~\cite{Takada:2013wfa}), we include a detailed derivation here to fix notation and to clarify the assumptions entering our analysis.

\subsection{Conventions}
We assume Gaussian initial conditions and a $\Lambda$CDM cosmology corresponding to the fiducial $\textsc{Quijote}$ simulations~\cite{Villaescusa-Navarro:2019bje}: $\Omega_m=0.3175$, $\Omega_{b,0}=0.049$, $h=0.6711$, $n_s=0.9624$, and $\sigma_8=0.834.$ 

We neglect redshift-space distortions and use $\delta(\bx)$ to denote a general 3D real-space density field at a fixed redshift $z$, which we typically omit for brevity. In this work, we focus exclusively on the matter density contrast, 
\begin{equation}
    \delta_m(\bx)\equiv\rho_m(\bx)/\bar{\rho}_m-1.
\end{equation}
We work in Fourier space, where the density field is related to its configuration-space counterpart by
\begin{equation}
    \delta({\bx})= \int \frac{d^3k}{(2\pi)^3}\,\delta(\vk)e^{-i\vk \cdotp \bx}\equiv \int_{\vk}\delta(\vk)e^{-i\vk\cdotp\bx}.
\end{equation}
The power spectrum, bispectrum, and trispectrum are defined by
\begin{align}
    \langle \delta(\vk_1)\delta(\vk_2)\rangle'_{c} &\equiv P(k_1)\,, \\
    \langle \delta(\vk_1) \delta(\vk_2) \delta(\vk_3) \rangle'_{c} &\equiv B(k_1, k_2,k_3)\,,\\
        \langle \delta(\vk_1) \delta(\vk_2) \delta(\vk_3)\delta(\vk_4) \rangle'_{c} &\equiv T(k_1, k_2,k_3,k_4, k_{12},k_{14})\,,
\end{align}
respectively, where $\langle \cdots \rangle'_c$ denotes the connected correlation function with the overall momentum-conserving delta function stripped off, and $\vk_{1\dots n}\equiv \vk_1+\dots+\vk_n$. The bispectrum is parameterized by the side lengths of three momentum vectors forming a triangle, while the trispectrum is parameterized by the four side lengths and two diagonals spanning a tetrahedron. We use $\delta_D^{(3)}$ and $\delta^{\rm K}_{k,k'}$ to denote the Dirac and Kronecker deltas, respectively.

As we show below, the power spectrum covariance is sensitive to specific ``soft" limits of the bispectrum and trispectrum, in which one momentum is much smaller than the others. In particular, we consider the squeezed bispectrum with $q\equiv k_1 \ll k_{\rm NL}$ and $k_2\simeq k_3\equiv k$, as well as the collapsed trispectrum with $q\equiv k_{12} \ll k_{\rm NL}$, $k_1\simeq k_2\equiv k$, and $k_3\simeq k_4\equiv k'$. Here, $k_{\rm NL}$ denotes the non-linear scale, so the soft mode is always in the linear regime. Conversely, the ``hard" modes $k$ and $k'$ can be in the non-linear regime, as long as $k,k'\gg q.$

Finally, since we typically work with binned correlation functions, we reserve $q$ and $k$ without superscripts for continuous wavenumbers, and denote binned quantities averaged over some bins $b$ and $b'$, with the corresponding binned wavenumber written as $k_{b}$ and $k_{b'}$, respectively.

\subsection{Consistency relations and large-scale structure correlation functions}\label{subsec:CRs}
\noindent Before discussing power spectrum covariance estimation, we briefly review how symmetries constrain the soft limits of large-scale structure correlation functions. These statements, commonly referred to as large-scale structure consistency relations (see, \emph{e.g.}, Refs.~\cite{Peloso:2013zw, Kehagias:2013yd, Creminelli:2004yq,Cheung:2007sv,Tanaka:2011aj,Creminelli:2012ed,Hinterbichler:2012nm,Assassi:2012zq,Kehagias:2012pd,Pajer:2013ana,Hinterbichler:2013dpa,Goldberger:2013rsa,Baldauf:2015xfa,Bravo:2017gct,Hui:2018cag}), fix the long-wavelength dependence of the squeezed bispectrum and collapsed trispectrum, and, hence, that of the power spectrum covariance.

We begin with the squeezed bispectrum, which is directly related to the super-sample covariance. The squeezed matter bispectrum\footnote{For concreteness, we focus on the matter field here, since we are interested in the matter power spectrum covariance; nevertheless, many of the results in this section can be straightforwardly generalized to biased tracers, such as galaxies and halos, which also obey large-scale structure consistency relations. } is given by the correlation of a long-wavelength mode, $\delta_{m,L}(\vq,z)$\footnote{We use the subscript $\delta_{m,L}$ to emphasize that this is a long-wavelength mode that is in the linear regime.} with the locally measured small-scale isotropic power spectrum, $P_m(k,z|\vx)|_{\delta_L}$. To proceed, we expand $P_m(k,z|\vx)|_{\delta_L}$ in terms of the long-wavelength background perturbation, 
\begin{align*}\label{eq: squeezed_bispectrum_integral}
    \begin{split}
        \lim_{q\ll k}\,B_m(q,k,z) &=\av{\delta_{m,L}(q) P_m(k\,|\,\vx)|_{\delta_L}}_c'\,, \\
        &=\av{\delta_{m,L}(q)\left(\int_{\vp} \, \frac{\partial P_m(k)}{\partial\,\delta_{m,L}(\vp)}\, \delta_{m,L}(q) \right)}_c'\,, \\
        &=\left(\frac{\partial P_m(k)}{\partial\,\delta_{m,L}(q)}\right) P_m^{\rm lin.}(q).
    \end{split}
\end{align*}
Here, ${\partial P_m(k)}/{\partial\,\delta_{L}(q)}$ represents the response of the small-scale isotropic power spectrum to a long-wavelength linear density perturbation. While we focus exclusively on the angle-averaged (isotropic) bispectrum and power spectrum, our results can be generalized to anisotropic power spectra and long-wavelength tidal fields, following, \emph{e.g.}, Refs.~\cite{Barreira:2017sqa,Chiang:2018mau,Masaki:2020drx}.

 In the absence of primordial non-Gaussianity and equivalence-principle-violating physics, the $q$-dependence of the response function takes the following form
\begin{equation}\label{eq:Response_def}
    \frac{\partial P_m(\vk)}{\partial\,\delta_{L}(\vq)}=a_0(k)+\mathcal{O}(q^2/k^2),
\end{equation}
where $a_0(k)$ encodes the isotropic response of the matter power spectrum response to a long-wavelength density perturbation.\footnote{For brevity, we absorb $P_{m}(k)$ in the definition of $a_0(k)$, hence $a_0(k)$ has units of volume. $a_0(k)$ is directly related to the isotropic response $\mathcal{R}_1(k)$ of Ref.~\cite{Barreira:2017sqa}, where $a_0(k)=\mathcal{R}_1(k)P_{m}(k)$. It is also closely related to the ``Gaussian bias'' $b_{\pi}$ of Ref.~\cite{Giri:2023mpg}.} Consequently, the squeezed matter bispectrum can be written as 
\begin{equation}\label{eq:squeezed_bispec_model}
    \lim_{q\ll k}B_m(q,k,z)=a_0(k)P^{\rm lin.}_m(q)\propto P^{\rm lin.}_m(q).
\end{equation}
Importantly, this result is fully non-perturbative, \emph{i.e.}, the long-wavelength dependence of the squeezed bispectrum must follow the linear matter power spectrum, regardless of the complexities of non-linear structure formation.\footnote{A potentially more intuitive way to understand this is to note that the locally measured small-scale power spectrum is itself a biased tracer of the long-wavelength density field. In the absence of equivalence-principle-violating physics or primordial non-Gaussianity, the cross power spectrum between any biased tracer and the underlying matter density must follow the linear matter power spectrum on large scales by symmetry~\cite{Desjacques:2016bnm}.} Indeed, the underlying rationale of Refs.~\cite{Goldstein:2022hgr, Giri:2023mpg, Goldstein:2023brb, Goldstein:2024bky, Goldstein:2025eyj, Zang:2025azh} was to use violations in this relationship as a probe of primordial non-Gaussianity in the non-linear regime.

We now turn to the collapsed matter trispectrum. In this limit, the trispectrum depends on the soft mode $q\equiv k_{12}$ and two external momenta $k\equiv k_1\simeq k_2$ and $k'\equiv k_3\simeq k_4$. The collapsed trispectrum describes the correlation between two locally measured small-scale power spectra, $\av{ P_{m}(k|\bx)|_{\delta_L} P_{m}(k'|\bx)|_{\delta_L}}'$, separated by a long-wavelength perturbation $q$ (see, \emph{e.g.}, Refs.~\cite{Lewis:2011au, Kenton:2016abp} and Fig.~\ref{fig:trispectrum_diagram}). This contains the $q$-dependent contribution
\begin{align}\label{eq:collapsed_T_no_contact}
    T_m(q,k,k')&\supset\frac{\partial P_m(k)}{\partial \delta_L(\vq)}\frac{\partial P_m(k')}{\partial \delta_L(\vq)}P_{m}^{\rm lin.}(q)\nonumber,\\
    &=a_0(k)a_0(k')P_m^{\rm lin.}(q).
\end{align}
Thus, in the absence of primordial non-Gaussianity and equivalence-principle-violating physics, the soft-mode dependence of the collapsed matter trispectrum is entirely fixed by the linear matter power spectrum. In addition to this $q$-dependent term, small-scale mode coupling can generate $q$-independent contributions to the trispectrum, which we denote by $\mathcal{T}_0(k,k')$. In total, the collapsed matter trispectrum can be written as
    \begin{equation}\label{eq:matter_trispectrum_theory}
     T_m(q,k,k')= a_0(k)a_0(k')P_m^{\rm lin.}(q)+\mathcal{T}_0(k,k').
    \end{equation}
The $q$-independent and $q$-dependent contributions to the collapsed trispectrum set the connected non-Gaussian and super-sample contributions to the power spectrum covariance, respectively.

\subsection{Binned power spectrum, bispectrum, and trispectrum estimation}\label{Sec:estimators}

Having discussed the theoretical models for the squeezed bispectrum and collapsed trispectrum, we now turn to practical estimators for the binned power spectrum, bispectrum, and trispectrum. Ultimately, we aim to relate the covariance of the estimated power spectrum to the estimated bispectrum and trispectrum. 

Consider a generic 3D density field, $\delta$, which could represent, \emph{e.g.}, the matter or galaxy density. An estimator for the binned power spectrum of $\delta$ in some bin $k_b$ is
\begin{equation}\label{eq:Pk_estimator}
    \hat{P}(k_b) = \frac{1}{N_{k_b}} \int\limits_{\vp_1, \vp_2\in k_b} \delta(\vp_1)\delta(\vp_2)\, (2\pi)^3 \delta_D^{(3)}(\vp_1+\vp_2),
\end{equation}
where $\int\limits_{\vp_1\in k_b}\equiv \int\limits_{\vp_1}\Theta_{k_b}(\vp_1)$ is shorthand for an integral over a bandpass filter $\Theta_{k_b}$ that selects modes in a bin, which we take to be an indicator function,\footnote{Throughout, we always assume an isotropic filter with $\Theta_{k_b}(p)^2 = \Theta_{k_b}(p)$.}
\begin{equation}\label{eq:fourier_filt}
    \Theta_{k_b}(p) = \begin{cases} 1, & p \in k_b\\ 0, & \text{otherwise},\end{cases}
\end{equation}
and $N_{k_b}$ is the number of Fourier modes in the bin, 
\begin{equation}
    N_{k_b} = V \int_{\vp} \Theta_{k_b}(p),
\end{equation}
where $V$ is the survey volume.
In practice, one can integrate over the delta function directly in Eq.~\eqref{eq:Pk_estimator}, yielding $\hat{P}(k_b) = N_{k_b}^{-1}\int_{\vp_1\in k_b}|\delta(\vp_1)|^2$. However, when writing out estimators for higher-order statistics, such as the bispectrum and trispectrum, it is useful to replace the delta function with its integral representation,
\begin{equation}
    \delta_{D}^{(3)}(\vk)=\frac{1}{(2\pi)^3}\int\, d^3x\,e^{-i\vk\cdotp \vx},
\end{equation}
yielding the following (equivalent) power spectrum estimator
\begin{equation}\label{eq:Pk_estimator_real}
    \hat{P}(k_b) =\frac{1}{N_{k_b}}\int\, d^3x\,\delta_{k_b}(\vx)^2, 
\end{equation}
where
\begin{equation}
    \delta_{k_b}(\vx) \equiv \int_{\vp}\,\Theta_{k_b}(\vp)\,\delta(\vp)\,e^{-i\vp\cdotp\vx}.
\end{equation}

Similarly, the binned bispectrum can be estimated using
\begin{align}\label{eq:bispectrum_estimator}
    \hat{B}(k_{b_1},k_{b_2}, k_{b_3}) &\propto \bigg(\prod_{i=1}^{3}\,\int\limits_{\vp_i\in k_{b_i}}\delta(\vp_i)\bigg)
    (2\pi)^3 \delta_D^{(3)}(\vp_{123}),\nonumber\\
    &=\int d^3x\,\delta_{k_{b_1}}(\vx)\,\delta_{k_{b_2}}(\vx)\,\delta_{k_{b_3}}(\vx),
\end{align}
where $\vp_{123} = \vp_1+\vp_2+\vp_3$. Here, we have omitted the normalization $N(k_{b_1}, k_{b_2}, k_{b_3})$, which counts the number of triangles in a bin configuration, and can be evaluated by replacing $\delta(\vp_i)\rightarrow 1$ in the first line of Eq.~\eqref{eq:bispectrum_estimator}. In the second line of Eq.~\eqref{eq:bispectrum_estimator}, we wrote the bispectrum estimator in a separable form that can be easily evaluated using Fast Fourier Transforms (FFTs).

A binned estimator for the total four-point function, which includes disconnected contributions, can be derived in a similar fashion. However, since the binned trispectrum estimator as a function of all six momenta is not separable, we use the estimator from Refs.~\cite{Coulton:2023oug, Goldstein:2024bky}, which integrates over one of the internal momenta, $k_{14}$. In Appendix~\ref{App:trispectrum_estimators}, we discuss this estimator in detail. Here, we focus on the binned trispectrum in the collapsed limit,  which is the relevant configuration for covariance estimation. Thus, we consider the trispectrum estimated in a soft-mode bin $q_b$ and two (possibly equal) hard-mode bins $k_{b}$ and $k_{b}'$. An estimator for the total four-point function in this limit is\footnote{We use $\delta_\vp\equiv \delta(\vp)$ for convenience here.}
\begin{widetext}
    \begin{align}
            \hat{T}_{\rm tot.}(q_b, k_b, k_b')&\propto \int\limits_{\vP}\Theta_{q_b}(P)\int\limits_{\vp_1,\dots, \vp_4}\bigg[\Theta_{k_b}(p_1)\,\Theta_{k_b}(p_2)\,\Theta_{k_b'}(p_3)\,\Theta_{k_b'}(p_4)\,\delta_{\vp_1}\delta_{\vp_2}\delta_{\vp_3}\delta_{\vp_4}
     (2\pi)^3\delta_D^{(3)}(\vp_{12}-\vP)(2\pi)^3\delta_D^{(3)}(\vp_{34}+\vP)\bigg]\nonumber,\\
     &\propto \int_{\vP}\Theta_{q_b}(P) \left( \int d^3x\,e^{-i\vP\cdotp \bx}   \delta_{k_b}(\vx)^2\right) \left( \int d^3y\,e^{i\vP\cdotp \by}  \delta_{k_b'}(\by)^2  \right),\label{eq:tot_trispectrum_with_sep_form}
    \end{align}
\end{widetext}
where the second line shows the separable form of the estimator, revealing the connection between the collapsed trispectrum estimator and the power spectrum estimator (Eq.~\eqref{eq:Pk_estimator_real}). For brevity, we have omitted the normalization in Eq.~\eqref{eq:tot_trispectrum_with_sep_form}, which counts the number of tetrahedra in a given momentum bin configuration and can be evaluated by replacing $\delta(\vp_i)\rightarrow 1$. We write out the full normalization in Appendix~\ref{App:trispectrum_estimators}, as it plays an important role in recovering the \emph{Gaussian} covariance from the total four-point function estimator.

 Eq.~\eqref{eq:tot_trispectrum_with_sep_form} includes disconnected contributions that must be subtracted to obtain an estimator for the trispectrum. We subtract disconnected contributions using the estimator from Ref.~\cite{Goldstein:2024bky}, which we briefly review here and detail further in Appendix~\ref{App:trispectrum_estimators}. In the limit of mild non-Gaussianity, the optimal trispectrum estimator is~\cite{Smith:2015uia, Shen:2024vft},
 \begin{equation}
     \hat{T}_{\rm opt}=\delta^4-6\,\delta^2\langle \delta^2\rangle+3\langle \delta^2\rangle^2.
 \end{equation}
  Consequently, a quasi-optimal estimator for the trispectrum can be constructed by replacing $\delta^4$ in the top line of Eq.~\eqref{eq:tot_trispectrum_with_sep_form} with $\delta^4-6\,\delta^2\langle \delta^2\rangle+3\langle \delta^2\rangle^2$, where $\langle \delta^2(p)\rangle'_c=P_{\rm fid}(p)$ is evaluated using some fiducial power spectrum. This estimator is sensitive to errors in the fiducial power spectrum at order $(\Delta P(k))^2$~\cite{Smith:2015uia, Shen:2024vft}. Hence, a $10\%$ error in the fiducial power spectrum biases the trispectrum by only $\sim 1\%$ (see Appendix~\ref{App:disc_convergence} as well Ref.~\cite{Goldstein:2024bky} for more details regarding this point). We note that this approach is analogous to the realization-dependent $N^0$ bias subtraction technique commonly used in CMB lensing reconstruction~\cite{Hanson:2010rp,Namikawa:2012pe}.

Finally, we note that one can construct a near-optimal, data-only disconnected estimator that does not require a fiducial power spectrum, of the form $\hat{T} \sim \delta^4 - 3\,\delta^2 \delta^2$ (see Ref.~\cite{Shen:2024vft} for a similar approach in the context of CMB lensing). We discuss this estimator in Appendix~\ref{App:trispectrum_estimators} and Appendix~\ref{App:disc_convergence}; however, we adopt the optimal estimator in the main text, as it is better suited for observational systematics, such as partial sky coverage, even though we neglect these in the current work.

\subsection{Power spectrum covariance}\label{Sec:Pk_cov_background}

\noindent We now connect the power spectrum covariance to the squeezed bispectrum and collapsed trispectrum. Although this derivation follows standard arguments in the literature~\cite{Takada:2013wfa, Li:2014sga, Barreira:2017kxd, Barreira:2017fjz}, we include it here to explicitly relate the power spectrum covariance to the trispectrum estimator defined in Eq.~\eqref{eq:tot_trispectrum_with_sep_form}. The power spectrum covariance in two momentum bins $k_b$ and $k_b'$ can be decomposed as follows
\begin{align}\label{eq:cov_def}
    \mathrm{Cov}_{k_b, k_b'} &\equiv \langle \hat{P}(k_b)\hat{P}(k_b')\rangle - \langle \hat{P}(k_b)\rangle \langle \hat{P}(k_b')\rangle \\
    &= \mathrm{Cov}^{\rm G}_{k_b,k_b'} + \mathrm{Cov}^{\rm NG}_{k_b,k_b'} \\
    &= \mathrm{Cov}^{\rm G}_{k_b,k_b'} + \mathrm{Cov}^{\rm cNG}_{k_b,k_b'} + \mathrm{Cov}^{\rm SSC}_{k_b,k_b'}.
\end{align}
Here, $\mathrm{Cov}^{\rm G}$ is the Gaussian contribution from the disconnected four-point function, which dominates on large scales where the trispectrum is negligible. On sufficiently small scales, the non-Gaussian contributions dominate. The non-Gaussian covariance is conventionally separated into two separate terms: the connected non-Gaussian term, $\mathrm{Cov}^{\rm cNG}$, arises from mode coupling within the survey, while the super-sample covariance, $\mathrm{Cov}^{\rm SSC}$, comes from coupling between small-scale modes inside the survey and large-scale modes outside of the survey. In this section, we derive the Gaussian and connected non-Gaussian covariance associated with the power spectrum estimator in Eq.~\eqref{eq:Pk_estimator}. We also discuss the super-sample covariance, although we defer a derivation of that result to Appendix~\ref{App:cov_derivation} because it relies on introducing a survey window function.

\subsubsection{Gaussian covariance}
The Gaussian covariance can be obtained from Eq.~\eqref{eq:cov_def} by assuming $\delta$ is a Gaussian random field and using Wick's theorem to simplify the four-point function, \emph{i.e.},
\begin{widetext}
\begin{align}\label{eq:Gaussian_cov_derivation}
    \langle \hat{P}(k_b)\hat{P}(k_b')\rangle_{\rm G}
    &= \frac{1}{N_{k_b}N_{k_b'}}\int\limits_{\vp_1\dots\vp_4}\Theta_{k_b}(p_1)\Theta_{k_b}(p_2)\Theta_{k_b'}(p_3)\Theta_{k_b'}(p_4)\nonumber\\
    &\quad\times
    (2\pi)^3\delta_D^{(3)}(\vp_{12})
    (2\pi)^3\delta_D^{(3)}(\vp_{34}) \Big[P(p_1)P(p_2) (2\pi)^3\delta_D^{(3)}(\vp_{12}) (2\pi)^3\delta_D^{(3)}(\vp_{34})+ \text{2 perm.}\Big],\nonumber\\[6pt]
    &= P(k_b)P(k_b') + \frac{2P(k_b)^2}{N_{k_b}}\delta^{\rm K}_{k_b,k_b'}.
\end{align}
\end{widetext}
Therefore, the Gaussian covariance is 
\begin{equation}
    \cov^{\rm G}_{k_b,k_b'}=\frac{2P^2(k_b)}{N_{k_b}}\delta^{\rm K}_{k_b,k_b'}.
\end{equation}
Notice that the Gaussian covariance is diagonal in $k$ and decreases with the number of Fourier modes. In Appendix~\ref{App:GRFs}, we derive the relationship between the Gaussian power spectrum covariance and the disconnected contributions to our trispectrum estimator and find that, up to geometric factors, they are equivalent. In the remainder of this work, we assume that the Gaussian covariance, and hence the power spectrum, is known with reasonable accuracy.

\subsubsection{Connected non-Gaussian covariance}

In the absence of a survey window function, the connected non-Gaussian contribution is given by the connected contribution to the four-point function, \emph{i.e.},
\begin{widetext}
\begin{align}\label{eq:cnG_cov_derivation}
     \mathrm{Cov}^{\rm cNG}_{k_b,k_b'}&=
    \frac{1}{N_{k_b}N_{k_b'}}\int\limits_{\vp_1\dots\vp_4}\Theta_{k_b}(p_1)\Theta_{k_b}(p_2)\Theta_{k_b'}(p_3)\Theta_{k_b'}(p_4)(2\pi)^3\delta_D^{(3)}(\vp_{12})(2\pi)^3\delta_D^{(3)}(\vp_{34})(2\pi)^3\delta_D(\vp_{1234})\,T(\vp_1,\vp_2,\vp_3,\vp_4),\nonumber\\
    &=\frac{1}{V}\left[  \frac{V^2}{N_{k_b}N_{k_b'}}\int\limits_{\vp,\vp'}\Theta_{k_b}(p)\Theta_{k_b'}(p')\,T(\vp,-\vp,\vp',-\vp')\right],
\end{align}
\end{widetext}
where the quantity in brackets is the ``bin-averaged trispectrum." Notice that the connected non-Gaussian covariance is sensitive to the \emph{exact} collapsed limit of the trispectrum where the internal momentum is precisely zero. In other words, the connected non-Gaussian covariance is directly set by the $q$-independent contribution to Eq.~\eqref{eq:matter_trispectrum_theory},\footnote{Eq.~\eqref{eq:cov_NG_soft_limit} also applies to biased tracers, provided $\mathcal{T}_0(k_b,k_b')$ is replaced by the $q$-independent contribution to the trispectrum of the tracer, which will typically include shot noise contributions. However, the collapsed trispectrum of biased tracers will contain additional $q$-dependent terms compared to the matter trispectrum model in Eq~.\eqref{eq:matter_trispectrum_theory}.} hence
\begin{equation}\label{eq:cov_NG_soft_limit}
     \cov^{{\rm cNG}}_{k_b,k_b'}=\frac{1}{V}\bar{\mathcal{T}}_0^{k_b,k_b'},
\end{equation}
where $\bar{\mathcal{T}}_0^{k_b,k_b'}$ denotes the average value of the collapsed trispectrum in bins $k_b$ and $k_b'$. In Appendix~\ref{app:exact_collapsed_estimator}, we show that the bracketed term in Eq.~\eqref{eq:cnG_cov_derivation} is equivalent to taking the $q\rightarrow 0$ limit of the connected contribution to the four-point function estimator in Eq.~\eqref{eq:tot_trispectrum_with_sep_form}.

\subsubsection{Super-sample covariance}
\noindent To derive the super-sample covariance, one needs to introduce a survey window function. We include this derivation in Appendix~\ref{App:cov_derivation}, and quote the main results here. The super-sample covariance can be written as~\cite{Takada:2013wfa}
\begin{equation}
    {\rm Cov}^{\rm SSC}_{k_b,k_b'}=\sigma_W^2\left(\frac{\partial P_m(k_b)}{\partial \delta_L}\right)\left(\frac{\partial P_m(k_b')}{\partial \delta_L}\right),
\end{equation}
where $\partial P_m(k_b)/\partial \delta_L$ is the response of the isotropic power spectrum to a long-wavelength background density perturbation, and $\sigma_W^2$ is the variance of the linear background density field over the survey volume, 
\begin{equation}\label{eq:sigma_Pk_def}
    \sigma_W^2=\frac{1}{V^2}\int_{\vk}|W(\vk)|^2P_m^{\rm lin}(k).
\end{equation}
Here, $W(\vk)$ is the Fourier transform of the survey window function and $V=\int d^3\bx\, W(\bx)$ is the survey volume. Using Eq.~\eqref{eq:Response_def}, we can relate the super-sample covariance to the response function governing the squeezed bispectrum,
\begin{equation}
{\rm Cov}^{\rm SSC}_{k_b,k_b'}= \bar{a}_0^{k_b}\bar{a}_0^{k_b'}\,\sigma_W^2,
\end{equation}
where $\bar{a}_0^{k_b}$ denotes the average value of the response function in bin $k_b.$

\section{Methodology}\label{Sec:methodology}

Before describing the details of our analysis, we provide a high-level overview of our approach for estimating the non-Gaussian power spectrum covariance using bispectrum and trispectrum measurements. First, we measure the power spectrum, squeezed bispectrum, and collapsed trispectrum of the field. Second, we fit these measurements using the non-perturbative squeezed bispectrum (Eq.\eqref{eq:squeezed_bispec_model}) and trispectrum (Eq.\eqref{eq:matter_trispectrum_theory}) models, treating $\bar{a}_0^{k_b}$, $\bar{a}_0^{k_b'}$, and $\bar{\mathcal{T}}^{k_b,k_b'}$ as free parameters. Finally, we map the resulting constraints on these parameters directly to constraints on the super-sample and connected non-Gaussian power spectrum covariance. As we will show, we can achieve percent-level constraints on these parameters, and thus on the non-Gaussian covariance, using on the order of ten $(1~{\rm Gpc}/h)^3$ simulations across a broad range of scales. 

\subsection{Simulations and measurements}

To validate our approach for estimating the non-Gaussian power spectrum covariance, we use the \textsc{Quijote} simulation suite \cite{Villaescusa-Navarro:2019bje}. The \textsc{Quijote} simulations are an extensive suite of $N$-body simulations across a range of cosmologies and initial conditions in a $(1~{\rm Gpc}/h)^3$ volume, making them an excellent testbed for covariance estimation. We analyze the $z=0$ snapshots, where the non-Gaussian covariance is the most significant. We assign the matter field to a $N_{\rm grid}=512^3$ mesh using the cloud-in-cell (CIC) assignment scheme implemented in \textsc{Pylians}~\cite{Pylians},\footnote{\href{https://pylians3.readthedocs.io/en/master/}{https://pylians3.readthedocs.io/en/master/}} and estimate all power spectra, bispectra, and trispectra with binned FFT-based estimators implemented in the \texttt{PNGolin} package.\footnote{\href{https://github.com/samgolds/PNGolin}{https://github.com/samgolds/PNGolin}}

To evaluate the efficacy of our method, we first need precise measurements of the ``true" non-Gaussian matter power spectrum covariance. To this end, we estimate the connected non-Gaussian covariance using the sample covariance of the matter power spectrum measured across all 15,000 realizations of the fiducial \textsc{Quijote} suite. Similarly, we estimate the super-sample covariance using the \textsc{Quijote} separate universe simulations, which consist of 1,000 realizations with a uniform background density perturbation $|\delta_b|=0.035$ (500 with an overdensity and 500 with an underdensity). The separate universe response functions need to account for the modified cosmology used in the separate universe simulations. For the power spectrum covariance, the appropriate response function is (see, \emph{e.g.}, Refs.~\cite{Li:2014sga, Coulton:2022qbc}),
\begin{equation}\label{eq:SSC_SU_cov}
    \frac{\partial P(k)}{\partial \delta_{\rm b}}\bigg\vert_{\delta_{\rm b}=0}=P(k)+\frac{\partial P_{\rm SU}(k)}{\partial \delta_{\rm b}}-\frac{1}{3}\frac{\partial P(k)}{\partial \log(k)}.
\end{equation}
We estimate $\partial P_{\rm SU}(k)/\partial \delta_b$ via finite differencing the matter power spectra estimated from the $\delta_b=\pm0.035$ simulations, and we compute $\partial P(k)/\partial\log k$ by differentiating a cubic-spline interpolation of the mean power spectrum measured from 500 fiducial \textsc{Quijote} simulations. In total, we use 1,500 simulations to estimate the super-sample covariance and 15,000 to estimate the connected non-Gaussian covariance. This is sufficient for percent-level estimates of the non-Gaussian power spectrum covariance, enabling a stringent test of our formalism.

\begin{figure*}[!t]
\centering
\includegraphics[width=0.99\linewidth]{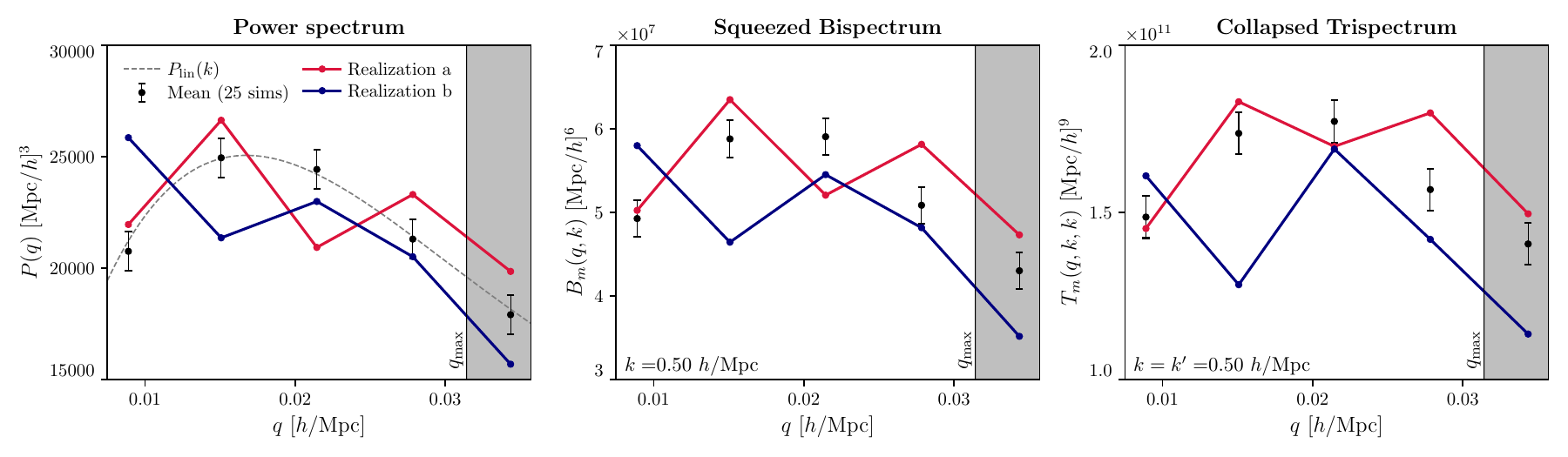}
\caption{Key summary statistics used in this work, measured from the \textsc{Quijote} $N$-body simulations at redshift $z=0$. From left to right, we show the large-scale power spectrum, the squeezed bispectrum, and the collapsed trispectrum of the matter field. Black points denote the mean over 25 realizations. For the squeezed bispectrum and collapsed trispectrum, the dependence on the long-wavelength mode $q$ is governed by $P(q)$. Crucially, all three statistics trace the same long-wavelength modes, so the squeezed bispectrum and collapsed trispectrum are highly correlated with the power spectrum on a realization-by-realization basis. This is illustrated by the blue and red points, which show measurements from two individual realizations. In this work, we exploit this sample variance cancellation when jointly fitting the power spectrum, bispectrum, and trispectrum to infer the small-scale power-spectrum covariance.} \label{fig:PBT_binned}
\end{figure*}

Having described the procedure for estimating the true non-Gaussian power spectrum covariance using conventional techniques, we now turn to the measurements needed in our new approach. We estimate the long-wavelength matter power spectrum $\hat{P}_m(q_b)$, squeezed bispectrum $\hat{B}_m(q_b,k_b)$, and collapsed trispectrum $\hat{T}_m(q_b,k_b,k_b')$ from 25 realizations of the fiducial \textsc{Quijote} simulations. We use five soft-mode bins spanning $0.006~h/{\rm Mpc}\simeq k_f \le q_b < 6k_f \simeq 0.037~h/{\rm Mpc}$ with bin width $\Delta q=k_f\simeq0.006~h/{\rm Mpc}$, where $k_f=2\pi/L_{\rm box}$ is the fundamental mode of the simulation. For the hard modes, we bin $k$ and $k'$ between $5k_f\simeq0.03~h/{\rm Mpc}$ and $120k_f\simeq0.75~h/{\rm Mpc}$ into twelve bins of width $\Delta k=10k_f\simeq0.06~h/{\rm Mpc}$. Since our goal is to demonstrate and assess the efficacy of our covariance estimation pipeline, we adopt a relatively coarse hard-mode binning scheme for computational reasons. Moreover, the non-Gaussian covariance is expected to vary smoothly as a function of $(k,k')$, so one could estimate it on a coarse grid and interpolate to a finer binning if needed.

We measure binned trispectrum configurations both with $k_b=k_b'$ and $k_b\neq k_b'$, corresponding to the diagonal and off-diagonal elements of the power spectrum covariance, respectively. When $k_b=k_b'$, the four-point function estimator includes disconnected contributions that source the Gaussian covariance. As shown in Appendix~\ref{App:GRFs}, the Gaussian and non-Gaussian contributions enter the estimator with different geometric weights, so the total covariance is not a simple rescaling of the total four-point function. To isolate the non-Gaussian power spectrum covariance, we subtract the disconnected contributions using the estimator described in Sec.~\ref{Sec:estimators}, which requires a fiducial power spectrum $P_{\rm fid}(k)$. We estimate $P_{\rm fid}(k)$ as the mean power spectrum measured from all 15,000 \textsc{Quijote} simulations, consistent with our assumption that the Gaussian covariance is known with sufficient accuracy. That being said, since the errors in our trispectrum estimator scale quadratically with the mismatch between $P_{\rm fid}(k)$ and the true power spectrum, we do not need anywhere near 15,000 simulations to estimate the disconnected terms. Indeed, in Appendix~\ref{App:disc_convergence}, we show that, for the survey volume and scale cuts considered here, our trispectrum estimator is converged using the power spectrum estimated from only a single realization. We also discuss trispectrum estimators that do not rely on an external power spectrum estimate in Appendix~\ref{App:trispectrum_estimators} and Appendix~\ref{App:disc_convergence}.

\subsection{Likelihood}\label{sec:likelihood}

Equipped with measurements of the power spectrum, squeezed bispectrum, and collapsed trispectrum, we now describe how we infer the non-Gaussian power spectrum covariance from these measurements. Consider estimating the off-diagonal covariance between two hard-mode bins $k_b$ and $k_b'$.\footnote{The diagonal case $k_b=k_b'$ follows identically, except that there is a single bispectrum $\hat B_m(q_b,k_b)$ and only two parameters, $\bar a_0^{k_b}$ and $\bar{\mathcal T}^0_{k_b,k_b}$.} For each soft-mode bin $q_{b_i}$, we define the data vector
\begin{equation}
    \mathbf{\hat{d}}_{b_i}\equiv\begin{pmatrix}
    \hat P_m(q_{b_i})\\
    \hat B_m(q_{b_i},k_b)\\
    \hat B_m(q_{b_i},k_b')\\
    \hat T_m(q_{b_i},k_b,k_b')
\end{pmatrix}.
\end{equation}
Our theory model for the data vector can be written entirely in terms of the linear matter power spectrum $P_m^{\rm lin}(q)$ (computed with \textsc{CLASS}~\cite{Lesgourgues:2011re}) and three free parameters, $\boldsymbol\theta_{k,k'} \equiv
(\bar a_0^{k_b},\;\bar a_0^{k_b'},\;\bar{\mathcal T}_0^{k_b,k_b'})$, \emph{i.e.},

\begin{equation}
\mathbf{d}^{\rm thr}_{b_i}(\boldsymbol\theta)\equiv
\begin{pmatrix}
P_m^{\rm lin}(q_{b_i})\\
\bar a_0^{k_b}P_m^{\rm lin}(q_{b_i})\\
\bar a_0^{k_b'}P_m^{\rm lin}(q_{b_i})\\
\bar a_0^{k_b}\bar a_0^{k_b'}P_m^{\rm lin}(q_{b_i})+\bar{\mathcal T}_0^{k_b,k_b'}
\end{pmatrix}. 
\end{equation}
Assuming uncorrelated $q_b$ bins,\footnote{In Appendix~\ref{App:likelihood}, we show that, over the range of scales considered here, the $q_b$ bins are approximately independent. Indeed, this lack of correlation is the main motivation for the present work.} the likelihood, up to an overall multiplicative constant, can be written as
\begin{equation} \label{eq:likelihood}
    \mathcal{L}_{k_b,k_b'}(\boldsymbol\theta) \propto \prod_{i} \frac{1}{\sqrt{\det{\mathbf C_{b_i}}}}\left[1+\frac{\Delta\mathbf{d}^{\!\top}\,{\mathbf C_{b_i}}^{-1}\,\Delta\mathbf{d}}{N_{\rm sim}-1}\right]^{-\frac{N_{\rm sim}}{2}}
\end{equation}
where $\Delta\mathbf{d} \equiv \mathbf{\hat{d}}_{b_i}-\mathbf{d}^{\rm thr}_{b_i}(\boldsymbol\theta)$ is the difference between the measured and theoretical data vectors, $\mathbf C_{b_i}$ is the $4\times4$ covariance matrix of $\hat{P}_m(q_{b_i})$, $\hat{B}_{m}(q_{b_i},k_b)$,$\hat{B}_{m}(q_{b_i},k_b')$, and $\hat{T}_{m}(q_{b_i},k_b,k_b')$, evaluated at fixed $q_i$, estimated using $N_{\rm sim}$ simulations, and the product runs over all $q$ bins. We use a multivariate $t$-distribution instead of a Gaussian likelihood because our covariance is estimated from a finite number of realizations~\cite{Sellentin:2015waz}. We discuss this in more detail in Appendix~\ref{App:likelihood}.

\begin{figure}[!t]
\centering
\includegraphics[width=0.99\linewidth]{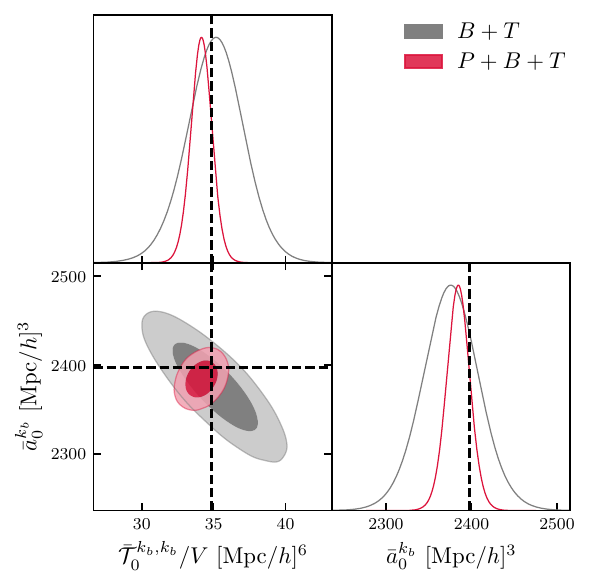}
\caption{Example marginalized posterior distributions for the parameters of our bispectrum and trispectrum models, obtained by fitting measurements from the \textsc{Quijote} simulations. The red contours show the 68\% and 95\% confidence regions from a joint likelihood analysis of the long-wavelength power spectrum $P_m(q)$, squeezed bispectrum $B_m(q,k)$, and collapsed trispectrum $T_m(q,k,k)$, averaged over 25 simulations for the small-scale bin centered at $k=0.50~h/{\rm Mpc}$ (corresponding to the black points in Fig.~\ref{fig:PBT_binned}). The black lines indicate the expected values inferred from independent estimates of the ``true" power spectrum covariance: $\bar{\mathcal{T}}^{k_b,k_b'}_{0}/V$, which determines the connected non-Gaussian covariance, is measured from the sample covariance of all 15,000 fiducial \textsc{Quijote} realizations, while $\bar{a}_0^{k}$, which controls the super-sample covariance, is obtained using the separate universe method with 1,500 simulations. Using only 25 realizations, we recover unbiased constraints on both parameters at percent-level precision. For reference, we include constraints obtained using only the squeezed bispectrum and collapsed trispectrum (gray), highlighting the power of sample variance cancellation in all three statistics (see Fig.~\ref{fig:PBT_binned}).} \label{fig:param_posterior}
\end{figure}

Since our method relies on the covariance of the long-wavelength power spectrum, squeezed bispectrum, and collapsed trispectrum, one might worry that we have replaced the original covariance-estimation challenge with a different one. In practice, however, only a small subset of this joint covariance is required. To determine the power spectrum covariance between a given pair of hard-mode bins $(k_b,k_b')$, we need at most the $4\times4$ covariance of $(\hat P(q_{b_i}), \hat B(q_{b_i},k_b), \hat B(q_{b_i},k_{b}'), \hat T(q_{b_i},k_b,k_b'))$ at fixed $q_{b_i}$. As demonstrated in Appendix~\ref{App:likelihood}, this covariance can be reliably estimated with $\mathcal{O}(10)$ simulations. Therefore, in the main text, we estimate $\mathbf{C}_i$ using only 25 realizations, such that our constraints on the power spectrum covariance are truly derived using only 25 realizations.\footnote{Technically, we use 15,000 simulations to estimate the fiducial power spectrum for computing the disconnected contributions to the trispectrum estimator; however, as shown in Appendix~\ref{App:disc_convergence}, we can obtain a converged estimate of the disconnected contributions using a single simulation.} Nevertheless, a more precise estimate of the full covariance would be necessary if one wished to compute the covariance of the inferred power spectrum covariance across many $(k,k')$ pairs. This higher-order quantity is not directly relevant for power-spectrum based cosmological parameter inference. Moreover, given the strong separation of scales underlying these observables, it may be possible to calculate the required covariances (semi-)analytically (see, e.g., Refs.~\cite{Biagetti:2021tua, Giri:2023mpg}). We leave a detailed investigation of approaches that circumvent the need for a trispectrum covariance estimate to future work.

Equipped with a likelihood and covariance, we sample from the parameter posterior using the \texttt{emcee} package~\cite{Foreman-Mackey:2012any} assuming wide uniform priors,
\begin{align*}
    -10^6~({\rm Mpc}/h)^3&<\bar{a}_0^{k}, \bar{a}_0^{k_b'} <10^6~({\rm Mpc}/h)^3,\\
    -10^{9}~({\rm Mpc}/h)^6&<\bar{\mathcal T}_0^{k_b,k_b'}/V<10^{9}~({\rm Mpc}/h)^6,
\end{align*}
where $V$ is the simulation volume. We assess convergence using the Gelman-Rubin statistic~\cite{gelman1992}, requiring $|R-1|<0.01$.

\begin{figure*}[!t]
\centering
\includegraphics[width=0.99\linewidth]{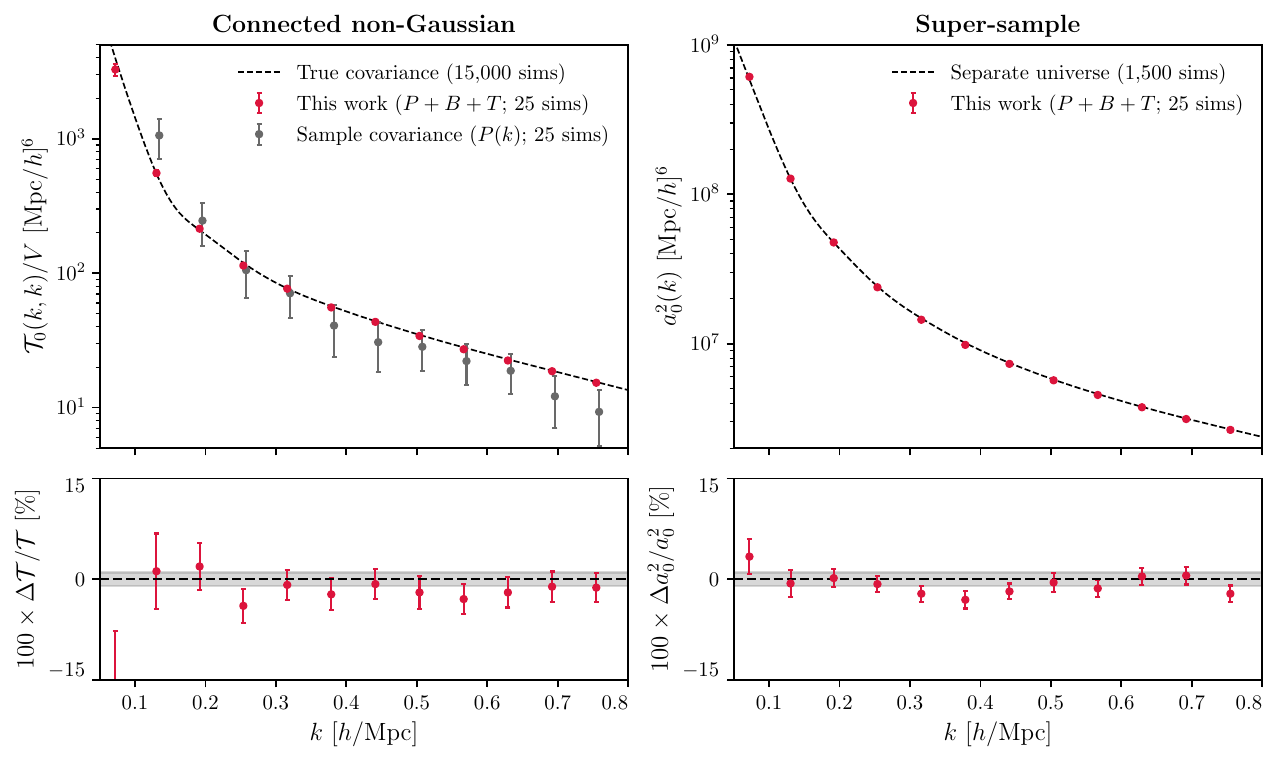}
\caption{Connected non-Gaussian (left) and super-sample (right) contributions to the matter power spectrum variance estimates. Red points show the results from a joint likelihood in the power spectrum, squeezed bispectrum, and collapsed trispectrum using 25 realizations. The dashed black curves indicate the ``true" covariance, computed from 15,000 simulations for the connected non-Gaussian covariance and 1,500 simulations for the super-sample covariance (500 fiducial plus 1,000 separate universe). The gray points in the top panel show the connected non-Gaussian covariance estimated using the sample covariance of 25 realizations, which is approximately 15$\times$ less precise than our estimator. For clarity, we do not rescale the super-sample contributions by $\sigma_W^2$ (see Eq.~\eqref{eq:sigma_Pk_def}), which depends on the survey volume. Note that data points across different $k$ bins are highly correlated because the non-Gaussian covariance is dense on these scales (see Fig.~\ref{fig:correlation_matrices}).}\label{fig:covariance_inference}
\end{figure*}

\section{Results}\label{Sec:results}
Fig.~\ref{fig:PBT_binned} shows the main summary statistics considered in this work: the long-wavelength power spectrum, the squeezed bispectrum, and the collapsed trispectrum. For the latter two, we show measurements averaged over a single hard-mode bin centered at $k_b=k_b'\approx 0.50~h/{\rm Mpc}$. The black points denote the mean over 25 realizations with 68\% uncertainties, while the red and blue lines show measurements from two individual realizations. The long-wavelength dependence of all statistics follows the linear matter power spectrum, consistent with the theoretical predictions in Sec.~\ref{subsec:CRs}. Furthermore, since all statistics trace the \emph{same} realization of the underlying long-wavelength modes, the different statistics are highly correlated on a realization-dependent level, as can be seen by comparing the red and blue curves across different panels. Crucially, this sensitivity to the same underlying long-wavelength modes enables a significant sample variance cancellation when the three statistics are combined in a joint likelihood.

Fig.~\ref{fig:param_posterior} shows the constraints on the model parameters $\bar{\mathcal{T}}^{k_b,k_b'}_{0}/V$ and $\bar{a}_0^{k_b}$, capturing the $q$-independent and $q$-dependent contributions to the collapsed trispectrum, respectively, derived directly from the measurements in Fig.~\ref{fig:PBT_binned}. The red contours denote the 68\% and 95\% confidence inferred by fitting the power spectrum, squeezed bispectrum, and collapsed trispectrum averaged over 25 realizations. Using just 25 realizations, we are able to constrain these parameters with percent-level uncertainties. Moreover, our constraints on $\bar{\mathcal{T}}^{k_b,k_b'}_{0}/V$ and $\bar{a}_0^{k_b}$ can be directly mapped onto the connected non-Gaussian and super-sample contributions to the matter power spectrum covariance at wavenumber $k$. This correspondence is illustrated by the dashed black lines, which show the reference (``true") covariance estimates obtained from 15,000 realizations for the connected non-Gaussian covariance and 1,500 realizations for the super-sample covariance. 

\begin{figure*}[!t]
\centering
\includegraphics[width=0.99\linewidth]{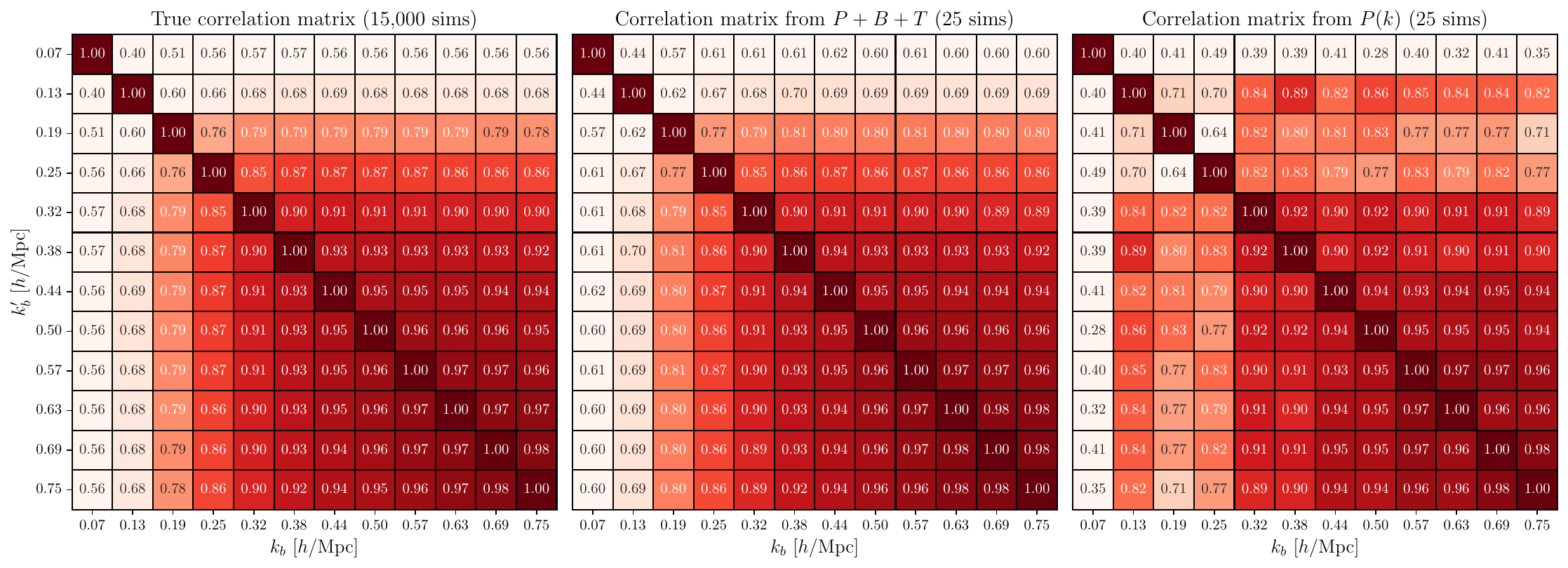}
\caption{Correlation matrices of the matter power spectrum covariance, showing Gaussian and connected non-Gaussian contributions. \emph{Left}: the ``true" correlation matrix from the sample covariance of 15,000 realizations. \emph{Middle}: the correlation matrix inferred by fitting the squeezed bispectrum and collapsed trispectrum using only 25 realizations, which reproduces the true correlations at the percent level. \emph{Right}: the sample correlation matrix from the power spectrum of 25 realizations, which is substantially noisier than our trispectrum-based estimator.} \label{fig:correlation_matrices}
\end{figure*}
To demonstrate the importance of sample variance cancellation, we repeat the analysis using only the squeezed bispectrum and collapsed trispectrum (gray contours). In this case, the constraints on $\bar{a}_0^{k_b}$ and $\bar{T}_0^{k_b,k_b}$ degrade by a factor of $\sim 2.5$ relative to the fiducial analysis, which includes information from the particular \emph{realization} of the long-wavelength power spectrum.

Having demonstrated that our method yields unbiased estimates of the non-Gaussian power spectrum covariance applied to a single $k_b$ bin, we now study its performance across a range of $k_b$ bins. Fig.~\ref{fig:covariance_inference} compares the constraints on the connected non-Gaussian (left) and super-sample (right)  matter power spectrum variance based on several methods considered in this work. For the connected non-Gaussian covariance, the dashed curve shows the true connected non-Gaussian contribution, estimated from the sample variance of 15,000 simulations. The gray points show the sample covariance estimated using just 25 simulations, with 68\% error bars estimated by bootstrap resampling across all 15,000 realizations. The red points show the mean and 68\% confidence interval on the connected non-Gaussian covariance, \emph{i.e.}, $\bar{\mathcal{T}}_{0}^{k_b,k_b'}/V$, derived from a joint analysis of the power spectrum, squeezed bispectrum, and collapsed trispectrum of these same 25 realizations. Using the bispectrum and trispectrum, we obtain an unbiased estimate of ${\rm cov}^{\rm cNG}$ at the percent level. The resulting uncertainty is roughly a factor of 15 smaller than that from the sample covariance of the power spectrum alone, implying that the standard approach would require about $225\times$ more simulations to achieve comparable precision. This gain is broadly consistent, albeit smaller than, the simple mode-counting expectation $\sqrt{\sum\limits_{q_i} N_{\rm modes}(q_i)}=\sqrt{484}= 22$. 

For the super-sample covariance, the red points show the constraints derived from the joint likelihood in the long-wavelength power spectrum, squeezed bispectrum, and collapsed trispectrum. Specifically, the red points denote the mean and 68\% confidence interval on $(\bar{a}_0^{k_b})^2$ defined in Sec.~\ref{sec:likelihood}. The black dashed curve shows the separate universe response $(\partial \log P_m(k)/\partial \delta_b|{\delta_b=0})^2$ from Eq.~\eqref{eq:SSC_SU_cov}, measured using 1,500 realizations (500 fiducial and 1,000 separate universe simulations). Using just 25 realizations, our method recovers an unbiased estimate of the separate universe response, and hence the super-sample covariance, within a few percent. Note that converting these response functions into the super-sample covariance requires multiplication by $\sigma_W^2$, which depends on the survey volume.

Having demonstrated that we can accurately estimate the diagonal elements of the matter power spectrum covariance, we now turn to the off-diagonal elements. Since the off-diagonal contributions from the super-sample covariance are fully determined by the diagonal terms, we focus here on the connected non-Gaussian covariance. Fig.~\ref{fig:correlation_matrices} compares the correlation matrices derived from several estimates of the connected non-Gaussian covariance. The left panel displays the ``true" correlation matrix from 15,000 realizations, while the middle panel shows our estimate using only 25 realizations. Our method reproduces the expected correlations to within a few percent across all scales, except for the smallest $k_b$ bin, for which the trispectrum is not sufficiently collapsed for our method to work. For comparison, the right panel shows the correlation matrix obtained from the sample covariance of the power spectrum using 25 realizations. The sample covariance is substantially noisier, with differences up to $20\%$.  Furthermore, while the \emph{correlation matrix} estimated from the sample covariance has the correct correlation structure above $k\gtrsim 0.3~h/{\rm Mpc}$, \emph{sample covariance} itself can still differ from the true power spectrum covariance by $\sim 30-40\%$, due to uncertainties in the power spectrum variance estimate (gray points in the top-left panel of Fig.~\ref{fig:covariance_inference}). We demonstrate this explicitly by plotting the full estimated covariance matrices in Appendix~\ref{App:full_covariance_matrix}. 

Overall, for the range of scales and simulation volumes considered here, our method recovers the connected non-Gaussian covariance to within a few percent using far fewer simulations than the sample covariance.

\section{Conclusions}\label{Sec:Conclusions}
Power spectrum covariance estimation is a central challenge of modern cosmological analyses. In this work, we developed a new method for estimating non-Gaussian contributions to the small-scale power spectrum covariance matrix by directly measuring and fitting the squeezed bispectrum and collapsed trispectrum of the density field. Intuitively, the collapsed trispectrum probes correlations between locally measured (small-scale) power spectra in the presence of a long-wavelength background mode. Therefore, each such background mode can be used to derive an approximately independent estimate of the small-scale power spectrum covariance. This is precisely the motivation behind internal covariance estimators, such as jackknife, which partition the survey into sub-volumes. However, the collapsed-trispectrum-based estimator presented here has several advantages over existing approaches because it naturally partitions the survey volume while self-consistently capturing the physical coupling between long and short-wavelength modes, enabling a clean separation between the connected non-Gaussian and super-sample covariance.

As a first test of our approach, we focused on estimating the 3D matter power spectrum covariance. Using the \textsc{Quijote} $N$-body simulations, we estimated (and separated) the super-sample and connected non-Gaussian contributions to the matter power spectrum covariance for $k\gtrsim0.15~h/{\rm Mpc}$ at redshift $z=0$ at the percent level using only 25 simulations. Our uncertainties on the connected non-Gaussian covariance are comparable to those one would achieve using the sample covariance of 5,000 simulations. Furthermore, our estimate of the super-sample covariance is derived directly from the simulations used for the connected non-Gaussian covariance, without the need for any separate universe simulations.

Beyond reducing the number of simulations needed for covariance estimation, our power spectrum covariance estimator opens up two possibilities. First, by comparing simulation-based covariances with the squeezed bispectrum and collapsed trispectrum measured directly from the survey, one could assess whether mock covariances accurately describe the data. Second, for a sufficiently large survey volume, one could estimate the power spectrum covariance directly from the data.

Despite these benefits, we note that our method relies on a separation of scales between the long- and short-wavelength modes in the survey. Concretely, we estimate the power spectrum covariance at wavenumbers $k$ and $k'$ using squeezed bispectrum and collapsed trispectrum estimates involving large-scale modes $q\leq q_{\rm max}\ll {\rm min}(k,k')$. Choosing $q_{\rm max}$ is a classic bias-variance tradeoff: increasing $q_{\rm max}$ improves errors, but if $q_{\rm max}$ is too large, two approximations break down. First, the bispectrum and trispectrum models in Eqs.~\eqref{eq:squeezed_bispec_model} and~\eqref{eq:matter_trispectrum_theory} become inapplicable, as the $(q/k)^2$ corrections are no longer negligible. Second, for sufficiently large $q$, the $q$ bins are not independent. In this work, we choose $q_{\rm max}=0.03~h/{\rm Mpc}$ (see detailed analysis in Appendix~\ref{App:likelihood}) and $k_{\rm min}\sim 5\times q_{\rm max} \approx 0.15~h/{\rm Mpc}$, as there is some evidence for bias at lower $k$ (see Fig.~\ref{fig:covariance_inference} and Appendix~\ref{App:full_covariance_matrix}). Although this restricts the applicability of our estimator on the largest scales, the modes with $k\lesssim 0.15~h/{\rm Mpc}$ are only mildly non-linear, and hence can be treated with analytic methods (see, \emph{e.g.}, Refs.~\cite{Mohammed:2014lja, Li:2018scc,Maus:2026wsb}).

To fully realize the potential of this method, we need to overcome several challenges. First, while we have focused on the real-space 3D matter field, large-scale structure surveys typically observe biased tracers in redshift space. Given the generality of large-scale structure consistency relations and the fact that the non-Gaussian power spectrum covariance is given by the collapsed trispectrum, it seems likely that the method presented here can be generalized to biased tracers in redshift space. In practice, however, biased tracers discretely sample the underlying matter field, leading to new contributions to the collapsed trispectrum that will need to be modeled to separate the connected non-Gaussian and super-sample covariances. It will also be important to generalize our framework to projected fields on the sphere, enabling applications to cosmic shear from weak lensing surveys as well as high-resolution measurements of the CMB temperature power spectrum, where non-Gaussian foregrounds contribute to the power spectrum covariance~\cite{AtacamaCosmologyTelescope:2025blo}.

We also note that our approach relies on a separation of scales between the long- and short-wavelength modes in the collapsed trispectrum. As a result, it is not applicable on scales comparable to the survey volume, where such a separation is not possible. Fortunately, for modern large-volume surveys, the non-Gaussian covariance is typically subdominant on these scales, hence one can use the Gaussian covariance, which can often be estimated analytically or with inexpensive mock simulations.

Beyond extending our methodology to more realistic large-scale structure tracers, it will be important to assess the impact of survey systematics. For example, a spatially varying selection function can induce large-scale modulations of the galaxy number density, hence a large collapsed trispectrum. Such effects are a challenge in large-scale constraints on $f_{\rm NL}$ from spectroscopic surveys~\cite{Chaussidon:2024qni}, and a careful treatment of these systematics will be essential to apply this method to realistic data since systematics do not obey the soft-limit structure assumed in our bispectrum and trispectrum models. Similarly, one would need to account for the survey window function. In this regard, we anticipate that recent progress in window-deconvolved power spectrum, bispectrum, and trispectrum estimators will be valuable~\cite{Philcox:2020vbm,Philcox:2021ukg,Philcox:2023uwe,Philcox:2023psd,Philcox:2024rqr}.

In principle, our framework could be extended to estimate the covariance of higher-order correlation functions, such as the bispectrum, although this would require developing estimators for five- and six-point correlation functions.\footnote{More generally, the covariance of any $N$-point function is set by soft limits of correlation functions up to order $2N$, and these soft limits satisfy consistency relations similar to the ones used in this work for the bispectrum and trispectrum.}

\acknowledgements

\noindent We thank Will Coulton for valuable comments on a draft of this manuscript. We thank Colin Hill, Nick Kokron, Mat Madhavacheril, Oliver Philcox, and Martin White for useful discussions. SG acknowledges support from NSF grant AST-2307727. The authors acknowledge the Texas Advanced Computing Center (TACC)\footnote{\href{http://www.tacc.utexas.edu}{http://www.tacc.utexas.edu}} at The University of Texas at Austin for providing computational resources that have contributed to the research results reported within this paper. We acknowledge computing resources from Columbia University's Shared Research Computing Facility project, which is supported by NIH Research Facility Improvement Grant 1G20RR030893-01, and associated funds from the New York State Empire State Development, Division of Science Technology and Innovation (NYSTAR) Contract C090171, both awarded April 15, 2010. M.M.  acknowledges support by the U.S. Department of Energy, Office of Science, Office of High Energy Physics under Award Number DE-SC0017647, by the National Science Foundation (NSF) under Grant Number 2307109 and 2509873 and by the Wisconsin Alumni Research Foundation (WARF). 

\bibliographystyle{apsrev4-1}
\bibliography{biblio}
\onecolumngrid

\clearpage
\appendix

\section{A review of 3D trispectrum estimators}\label{App:trispectrum_estimators}

\noindent In this appendix, we provide additional details regarding the three-dimensional (3D) trispectrum estimator used in this work, based on Refs.~\cite{Coulton:2023oug, Goldstein:2024bky} and implemented in the \texttt{PNGolin} package.\footnote{\href{https://github.com/samgolds/PNGolin}{https://github.com/samgolds/PNGolin}} Our goal is to derive a computationally efficient FFT-based estimator for the collapsed trispectrum and to connect this estimator to the non-Gaussian power spectrum covariance.

As in the main text, we parameterize the trispectrum in terms of four external momenta bins, $k_{b_1}$, $k_{b_2}$, $k_{b_3}$, and $k_{b_4}$, and two internal momenta bins, $Q_b\equiv k_{b_{12}}$ and $Q_b'\equiv k_{b_{14}}$, where $\Theta_{k_b}(p)$ is a bandpass filter that selects modes within bin $k_b.$ The binned trispectrum estimator as a function of all six momentum magnitudes is not separable; hence, we construct an estimator that is integrated over $Q_b'$. 

This appendix is divided into two sections. In the first section, we derive the estimator for a non-zero internal mode $Q_b$, which is the case used in this work. In the second section, we take the exact collapsed limit of the estimator ($Q_{b}=0$), and show that our estimator reduces to the standard sample covariance estimator of the power spectrum covariance.

\subsection{Collapsed trispectrum estimator with $Q_b\neq 0$}\label{app:trispectrum_estimator_nonzero_q}
\noindent  An estimator for the total four-point function, \emph{i.e.}, including disconnected contributions, is
\begin{align}\label{eq:total_estimator_Tk}
    \hat{T}_{\rm tot.}(k_{b_1}, k_{b_2}, k_{b_3}, k_{b_4}, Q_b)\propto \int\limits_{\vP}\Theta_{Q_b}(P)\int\limits_{\vp_1,\dots, \vp_4}\bigg[&\Theta_{k_{b_1}}(p_1)\Theta_{k_{b_2}}(p_2)\Theta_{k_{b_3}}(p_3)\Theta_{k_{b_4}}(p_4)\,\delta_{\vp_1}\delta_{\vp_2}\delta_{\vp_3}\delta_{\vp_4}\\
    & (2\pi)^3\delta_D^{(3)}(\vp_{12}-\vP)(2\pi)^3\delta_D^{(3)}(\vp_{34}+\vP)\bigg]\nonumber,
\end{align}
The normalization of the estimator can be computed by re-evaluating the integral with all density fields set to $1$, \emph{i.e.},
\begin{align}\label{eq:normalization_Tk}
N_{k_{b_1}, k_{b_2}, k_{b_3}, k_{b_4}, Q_b}= \int\limits_{\vP}\Theta_{Q_b}(P)\int\limits_{\vp_1}\Theta_{k_{b_1}}(p_1)\Theta_{k_{b_2}}(|\vP-\vp_1|)\int\limits_{\vp_3}\Theta_{k_{b_3}}(p_3)\Theta_{k_{b_4}}(|\vP+\vp_3|),
\end{align}
which counts the number of tetrahedra in a set of bins.

Eq.~\eqref{eq:total_estimator_Tk} can be efficiently evaluated using Fast Fourier Transforms (FFTs) by rewriting the Dirac delta functions as exponential integrals, yielding
\begin{equation}\label{eq:full_4pcf_estimator}
  \hat{T}_{\rm tot.}(k_{b_1}, k_{b_2}, k_{b_3}, k_{b_4}, Q_b)\propto \int_{\vP}\Theta_{Q_b}(P) \left( \int d^3x\,e^{-i\vP\cdotp \bx} \delta_{{k_{b_1}}}(\bx) \delta_{{k_{b_2}}}(\bx) \right) \left( \int d^3y\,e^{i\vP\cdotp \by} \delta_{{k_{b_3}}}(\by) \delta_{{k_{b_4}}}(\by) \right),
\end{equation}
where $ \delta_{{k_b}}(\bx)\equiv \int_{\vp}\Theta_{k_b}(p)\delta_{\vp}e^{i\vp\cdotp\bx}$ is the inverse Fourier transform of the filtered density field.

In order to isolate the connected contribution to Eq.~\eqref{eq:total_estimator_Tk}, \emph{i.e.}, the trispectrum, we need to subtract disconnected contributions from Eq.~\eqref{eq:total_estimator_Tk}. In the limit of mild non-Gaussianity, the optimal estimator can be obtained
by replacing $\delta^4 \rightarrow \delta^4-6\,\delta^2\langle \delta^2\rangle+3\langle \delta^2\rangle^2$ in Eq.~\eqref{eq:total_estimator_Tk}~\cite{Smith:2015uia, Shen:2024vft}. The $3\langle \delta^2\rangle^2$ term corresponds to a \emph{realization-independent} disconnected contribution,
\begin{equation}\label{eq:disc_real_indep}
    \hat{T}_{\rm disc.}^{\,3\langle \delta^2 \rangle^2} \propto \int\limits_{\vP}\Theta_{Q_b}(P)\left( \prod\limits_{i=1}^4\int\limits_{\vp_i}\Theta_{k_{b_i}}(p_i)\right) \left[ P_{\rm fid}(p_1)P_{\rm fid}(p_3)\,\delta_D^{(3)}(\vp_{12})\,\delta_D^{(3)}(\vp_{34})+2~{\rm perm.}\right]
     \delta_D^{(3)}(\vp_{12}-\vP)\delta_D^{(3)}(\vp_{34}+\vP),
\end{equation}
where $P_{\rm fid}(p_i)$ is some model for the power spectrum of $\delta$. Assuming the internal mode is non-zero,\footnote{See Appendix~\ref{app:exact_collapsed_estimator} for the estimator in the exact collapsed limit.} the first term in Eq.~\eqref{eq:disc_real_indep} is zero, and the two remaining permutations are
\begin{align*}
 \hat{T}_{\rm disc.}^{\,3\langle \delta^2 \rangle^2}&\propto \int\limits_{\vP}\Theta_{Q_b}(P)\left( \prod\limits_{i=1}^4\int\limits_{\vp_i}\Theta_{{k_{b_i}}}(p_i)\right) \left[ P_{\rm fid}(p_1)P_{\rm fid}(p_2)\,\delta_D^{(3)}(\vp_{13})\,\delta_D^{(3)}(\vp_{24})\right]
     \delta_D^{(3)}(\vp_{12}-\vP)\delta_D^{(3)}(\vp_{34}+\vP)+(3\leftrightarrow 4), \\
     & \propto \int_{\vP}\Theta_{Q_b}(P) \left( \int_{\vp_1}\Theta_{k_{b_1}}(p_1)\Theta_{k_{b_3}}(p_3)P_{\rm fid}(p_1)\right) \left( \int_{\vp_2}\Theta_{k_{b_2}}(p_2)\Theta_{k_{b_4}}(p_4)P_{\rm fid}(p_2)\right)\left[(2\pi)^3\delta_D^{(3)}(\vp_{12}-\vP)\right]^2+(3\leftrightarrow 4).
\end{align*}
To proceed, we replace the squared delta function. $((2\pi)^3\delta^{(3)}_D(\vk))^2=V\times (2\pi)^3\delta^{(3)}_D(\vk)$ (see, \emph{e.g.},~\cite{Gualdi:2020eag}), and express the remaining delta function as an exponential integral, yielding
\begin{equation}\label{eq:T_disc_3_PP}
\hat{T}_{\rm disc.}^{\,3\langle \delta^2\rangle^2}=\frac{V}{ N_{k_{b_1}, k_{b_2}, k_{b_3}, k_{b_4}, Q_b}}\int_{\vP}\Theta_{Q_b}(P)  \int d^3x\, e^{-i\vP\cdotp \bx}\left[ F^{P_{\rm fid}}_{13}(\bx)F^{P_{\rm fid}}_{24}(\bx)+F^{P_{\rm fid}}_{14}(\bx)F^{P_{\rm fid}}_{23}(\bx)\right], 
\end{equation}
where
\begin{equation}
    F^{P_{\rm fid}}_{ij}(\bx)\equiv \int_\vp \Theta_{k_{b_i}}(p)\Theta_{k_{b_j}}(p)P_{\rm fid}(p)e^{i\vp\cdotp\bm{x}}.
\end{equation}

Following the same reasoning used to derive the 3$\langle \delta^2\rangle^2$ term, the \emph{realization-dependent} $6\,\delta^2\langle \delta^2\rangle$ term can be written as
\begin{align}\label{eq:T_disc_6_deltaP}
    \hat{T}_{\rm disc.}^{6\,\delta^2\langle \delta^2\rangle}=\frac{V}{ N_{k_{b_1}, k_{b_2}, k_{b_3}, k_{b_4}, Q_b}}\int_{\vP}\Theta_{Q_b}(P)  \int d^3x\, e^{-i\vP\cdotp \bx}[& F^{P_{\rm fid}}_{13}(\bx)F^\delta_{24}(\bx)+F^\delta_{13}(\bx)F^{P_{\rm fid}}_{24}(\bx)\notag\\
    &+F^{P_{\rm fid}}_{14}(\bx)F^\delta_{23}(\bx)+F^\delta_{14}(\bx)F^{P_{\rm fid}}_{23}(\bx)],
\end{align}
where
\begin{equation}
    F^\delta_{ij}(\bx)\equiv \int_\vk \Theta_{k_{b_i}}(p)\Theta_{k_{b_j}}(p)|\delta_{\vp}|^2e^{i\vk\cdotp\bm{x}}.
\end{equation}
To summarize, a quasi-optimal estimator for the binned trispectrum is $\hat{T}_{\rm opt}= \hat{T}_{\rm tot.}-\hat{T}_{\rm disc.}^{6\,\delta^2\langle \delta^2\rangle}+\hat{T}_{\rm disc.}^{\,3\langle \delta^2 \rangle^2}$, where each term can be written in terms of FFTs using Eqs. \eqref{eq:total_estimator_Tk}, ~\eqref{eq:T_disc_3_PP}, and \eqref{eq:T_disc_6_deltaP}. 

Finally, it is possible to construct a data-only trispectrum estimator $\hat{T}=\hat{T}_{\rm tot.}-T_{\rm disc}^{3\delta^2\delta^2}$ where
\begin{equation}\label{eq:Tdisc_data_only}
    \hat{T}_{\rm disc.}^{\,3 \delta^2\delta^2}=\frac{V}{ N_{k_{b_1}, k_{b_2}, k_{b_3}, k_{b_4}, Q_b}}\int_{\vP}\Theta_{Q_b}(P)  \int d^3x\, e^{-i\vP\cdotp \bx}\left[ F^{\delta}_{13}(\bx)F^{\delta}_{24}(\bx)+F^{\delta}_{14}(\bx)F^{\delta}_{23}(\bx)\right].
\end{equation}
We test this data-only estimator in Appendix~\ref{App:disc_convergence}.

\subsection{Exact collapsed limit estimator ($Q_b=0$)}\label{app:exact_collapsed_estimator}

In the exact collapsed limit, the estimator can be derived by setting $\vP=0$ and taking $k_b \equiv k_{b_1}=k_{b_2}$ and $k_b' \equiv k_{b_3}=k_{b_4}$ in Eq.~\eqref{eq:total_estimator_Tk}.
\begin{align}\label{eq:total_estimator_Tk_fully_collapsed}
    \hat{T}_{\rm tot.}(k_b,k_b')&\propto \int\limits_{\vp_1,\dots, \vp_4}\bigg[\Theta_{k_b}(p_1)\Theta_{k_b}(p_2)\Theta_{k'_b}(p_3)\Theta_{k'_b}(p_4)\,\delta_{\vp_1}\delta_{\vp_2}\delta_{\vp_3}\delta_{\vp_4}(2\pi)^3\delta_D^{(3)}(\vp_{12})(2\pi)^3\delta_D^{(3)}(\vp_{34})\bigg]\nonumber,\\
    &=  \left[ \int d^3x\, (\delta_{k_b}(\bx))^2 \right] \times \left[ \int d^3y\, (\delta_{k_b'}(\by))^2 \right].
\end{align}
Therefore, in the exact collapsed limit, our trispectrum estimator is equivalent, up to an overall normalization, to the product of two power spectrum estimators. Thus, the collapsed trispectrum estimator directly corresponds to the connected non-Gaussian power spectrum covariance.

We can derive the disconnected contributions to the estimator using the same approach as the $q\neq 0$ estimator. The realization-independent disconnected contribution is
\begin{equation}
    \hat{T}_{\rm disc.}^{\,3\langle \delta^2 \rangle^2}\propto \left(\frac{V}{(2\pi)^3}\right)^2    \left[\,\int\limits_{\vp_1}\Theta_{k_b}(p_1) P_{\rm fid}(p_1)\right] \times \left[\,\int\limits_{\vp_3}\,\Theta_{k_b'}(p_3) P_{\rm fid}(p_3)\right]+\frac{2V}{(2\pi)^3}\int_{\vp_1} \Theta_{k_b}(p_1)\Theta_{k_b'}(p_1)P_{\rm fid}^2(p_1),
\end{equation}
Similarly, the realization-dependent disconnected estimator is 
\begin{align}
    \hat{T}_{\rm disc.}^{\,6\delta^2\langle \delta^2 \rangle}\propto &\left(\frac{V}{(2\pi)^3}\right)^2    \left\{\left[\,\int\limits_{\vp_1}\Theta_{k_b}(p_1) |\delta_{\vp_1}^2|\right] \times \left[\,\int\limits_{\vp_3}\,\Theta_{k_b'}(p_3) P(p_3)\right]+\left[\,\int\limits_{\vp_1}\Theta_{k_b}(p_1) P_{\rm fid}(p_1)\right] \times \left[\,\int\limits_{\vp_3}\,\Theta_{k_b'}(p_3)| \delta_{\vp_3}^2|\right]\right\}\nonumber\\
    &+\frac{4V}{(2\pi)^3}\int_{\vp_1} \Theta_{k_b}(p_1)\Theta_{k_b'}(p_1)P_{\rm fid}(p_1)|\delta_{\vp_1}^2|.
\end{align}
The data-only disconnected estimator can be derived as in Eq.~\eqref{eq:Tdisc_data_only}.

We now show that this estimator is equivalent to the connected non-Gaussian power spectrum covariance. Taking the expectation value of this estimator and retaining only the connected contribution, we find
\begin{align}
\hat{T}_{\rm conn}(k_b,k_b')&\propto \int\limits_{\vp_1,\dots, \vp_4}\bigg[\Theta_{k_b}(p_1)\,\Theta_{k_b}(p_2)\,\Theta_{k_b'}(p_3)\,\Theta_{k_b'}(p_4)\,T(\vp_1,\vp_2,\vp_3,\vp_4)(2\pi)^3\delta_D^{(3)}(\vp_{12})(2\pi)^3\delta_D^{(3)}(\vp_{34})(2\pi)^3\delta_D(\vp_{1234})\bigg]\nonumber,\\
&\propto V\int\limits_{\vp,\vp'}\Theta_{k_b}(p)\,\Theta_{k_b'}(p')\,T(\vp,-\vp,\vp',-\vp').
\end{align}
To relate this to the connected non-Gaussian power spectrum covariance (Eq.~\eqref{eq:cnG_cov_derivation}), we need to write out the normalization of the estimator,
\begin{align}
N(k_b,k_b')&= V\int\limits_{\vp_1,\dots, \vp_4}\bigg[\Theta_{k_b}(p_1)\,\Theta_{k_b}(p_2)\,\Theta_{k_b'}(p_3)\,\Theta_{k_b'}(p_4)\,(2\pi)^3\delta_D^{(3)}(\vp_{12})(2\pi)^3\delta_D^{(3)}(\vp_{34})\bigg]\nonumber,\\
&= V\left[\int_{\vp}\Theta_{k_b}(p)\right]\times  \left[\int_{\vp'}\Theta_{k_b}(p)\right]=V\left(\frac{N_{k_b}}{V}\right)\left(\frac{N_{k_b'}}{V}\right),
\end{align}
where $N_{k_b}=V\int_\vp \Theta_{k_b}(p)$ is the number of Fourier modes in the bin.

In conclusion, the connected contribution to the exact collapsed limit estimator is 
\begin{equation}
    \hat{T}_{\rm conn}(k_b,k_b')=\frac{V^2}{N_{k_b}N_{k_b'}}\int\limits_{\vp,\vp'}\Theta_{k_b}(p)\,\Theta_{k_b'}(p')\,T(\vp,-\vp,\vp',-\vp'),
\end{equation}
exactly matching the bracketed term in Eq.~\eqref{eq:cnG_cov_derivation}.

\clearpage
\pagebreak
\section{Derivation of the power spectrum covariance with a survey window}\label{App:cov_derivation}

In this appendix, we generalize the power spectrum covariance derivation from Sec.~\ref{Sec:Pk_cov_background} to include a survey window function, which is necessary to isolate the super-sample covariance (see, \emph{e.g.}, Refs.~\cite{Takada:2013wfa, Li:2014sga, Barreira:2017kxd, Barreira:2017fjz} for similar derivations). Let $W(\bx)$ denote the survey window, which is one inside the survey and zero outside. The windowed field is $\delta_W(\bx)\equiv W(\bx)\delta(\bx)$, hence, the windowed function convolves the density field,
\begin{equation}
    \delta_W(\vk) = \int_{\vp} W(\vp) \delta(\vk-\vp).
\end{equation} 

The power spectrum of the windowed field can be estimated as
\begin{equation}\label{eq:Pk_estimator_windowed}
    \hat{P}(k_b) = \frac{1}{N_{k_b}} \int_{\vp_1, \vp_2} \Theta_{k_b}(p_1) \Theta_{k_b}(p_2) \, \delta_W(\vp_1)\delta_W(\vp_2)\, (2\pi)^3 \delta_D^{(3)}(\vp_{12}).
\end{equation}
Here, $\Theta_{k_i}$ is a bandpass filter that selects Fourier modes, which we take to be an indicator function (Eq.~\eqref{eq:fourier_filt}), and $N_{k_b} = V \int_{\vp} \Theta_{k_b}(p)$ is the number of Fourier modes in bin $k_b$, where $V = \int d^3x\, W(\bx)$ is the survey volume.

The expectation value of the windowed power spectrum estimator is
\begin{equation}
    \langle \hat{P}(k_b)\rangle = \frac{1}{N_{k_b}} \int_{\vp} \Theta_{k_b}(p) \int_\vq W(\vp-\vq) P(q).
\end{equation}
Hence, Eq.~\eqref{eq:Pk_estimator_windowed} is a biased estimator of the true power spectrum on scales comparable to the survey volume. In practice, one must account for this bias by either directly modeling the convolution when computing theory predictions or explicitly deconvolving the window function (see, \emph{e.g.}, Refs~\cite{Tegmark:1997yq, Hamilton:2005kz, Hamilton:2005ma, Philcox:2020vbm}). In the main text, we analyze $N$-body simulations with periodic boundary conditions, for which $\hat{P}(k)$ is an unbiased estimator of the true power spectrum. Nevertheless, as we discuss in this appendix, the window function plays a crucial role in the classification of contributions to the non-Gaussian power spectrum covariance. 

Using the definition of the non-Gaussian power spectrum covariance (Eq.~\eqref{eq:cov_def}) and the windowed power spectrum estimator (Eq.~\eqref{eq:Pk_estimator_windowed}), the non-Gaussian covariance of the windowed power spectrum is
\begin{align}\label{eq:cov_NG_windowed}
      \mathrm{Cov}^{\rm NG}_{k_b,k_b'}&=\frac{1}{N_{k_b}N_{k_b'}}\int\limits_{\vp,\vp'}\Theta_{k_b}(p)\Theta_{k'_b}(p')\int\limits_{\vq_1\dots \vq_4}\left(\prod_{i=1}^{4}W(\vq_i)\right)T(\vp-\vq_1,-\vp-\vq_2,\vp'-\vq_3,-\vp'-\vq_4)(2\pi)^3\delta_D(\vq_{1234}).
\end{align}
We now consider the connected non-Gaussian and super-sample contributions to Eq.~\eqref{eq:cov_NG_windowed} separately.

\subsection{Connected non-Gaussian covariance}

The connected non-Gaussian covariance, $\cov^{\rm cNG}_{k_b,k_b'}$, arises from small-scale mode coupling within the survey volume. Since the trispectrum varies on scales $k\gg k_{\rm f}\sim V^{-1/3}$, it can be pulled outside the window function integrals, giving
\begin{align}\label{eq:cov_cNG_windowed}
      \mathrm{Cov}^{\rm cNG}_{k_b,k_b'}&=\frac{1}{N_{k_b}N_{k_b'}}\int\limits_{\vp,\vp'}\Theta_{k_b}(p)\Theta_{k'_b}(p')T(\vp,-\vp,\vp',-\vp')\times \left[\,\int\limits_{\vq_1\dots \vq_4}\left(\prod_{i=1}^{4}W(\vq_i)\right)(2\pi)^3\delta_D(\vq_{1234})\right],\\
      &= \frac{1}{V}\left[ \frac{V^2}{N_{k_b}N_{k_b'}}\int\limits_{\vp,\vp'}\Theta_{k_b}(p)\Theta_{k_b'}(p')T(\vp,-\vp,\vp',-\vp')\right],
\end{align}
consistent with Eq.~\eqref{eq:cnG_cov_derivation} in the main text. To obtain this result, we used the identity\footnote{The final equality here only holds for binary window functions, \emph{i.e.}, $W(\bx)\in \{0,1\}$. For more complicated window functions, one needs to explicitly keep track of powers of the survey window.}
\begin{equation}
    \int\limits_{\vq_1\dots \vq_n}\left(\prod_{i=1}^{n}W(\vq_i) \right)(2\pi)^3\delta_D(\vq_{1\dots n})=\int d^3x\,W^n(\bx)=V.
\end{equation}

\subsection{Super-sample covariance}

\noindent The super-sample covariance refers to the covariance sourced by the mode coupling of modes within the survey volume to modes that are larger than the survey volume. To derive this contribution starting with Eq.~\eqref{eq:cov_NG_windowed}, we make the change of variables $\vu\equiv \vp-\vq_1$ and $\vv\equiv \vp'-\vq_3$
\begin{align}\label{eq:cov_NG_windowed}
      \mathrm{Cov}^{\rm NG}_{k_b,k_b'}&=\frac{1}{N_{k_b}N_{k_b'}}\int\limits_{\vu,\vv}\int\limits_{\mathbf{q}_1\dots \mathbf{q}_4}\Theta_{k_b}(\mathbf{|\vu+\vq_1|})\Theta_{k_b'}(\mathbf{|\vv+\vq_3|})\left(\prod_{i=1}^{4}W(\mathbf{q}_i)\right)T(\mathbf{u},-\mathbf{u}-\mathbf{q}_{12},\mathbf{v},-\mathbf{v}-\mathbf{q}_{34})(2\pi)^3\delta_D(\mathbf{q}_{1234}),\nonumber\\
      &\approx \frac{1}{N_{k_b}N_{k_b'}}\int\limits_{\vu,\vv}\Theta_{k_b}(u)\Theta_{k_b'}(v)\int\limits_{\mathbf{q}_1\dots \mathbf{q}_4}\left(\prod_{i=1}^{4}W(\mathbf{q}_i)\right)T(\mathbf{u},-\mathbf{u}-\mathbf{q}_{12},\mathbf{v},-\mathbf{v}-\mathbf{q}_{34})(2\pi)^3\delta_D(\mathbf{q}_{1234}),
\end{align}
where we assumed that $k_b$ and $k_b'$ are much smaller than the survey volume. The super-sample covariance is sourced by the $q$-dependent contributions to the collapsed trispectrum (Eq.~\eqref{eq:collapsed_T_no_contact}), hence
\begin{align}
\mathrm{Cov}^{\rm NG}_{k_b,k_b'}&=\frac{1}{N_{k_b}N_{k_b'}}\int\limits_{\vu,\vv}\Theta_{k_b}(u)\Theta_{k_b'}(v)a_0(u)a_0(v)\int\limits_{\mathbf{q}_1\dots \mathbf{q}_4}\left(\prod_{i=1}^{4}W(\mathbf{q}_i)\right)P_m^{\rm lin.}(q_{12})(2\pi)^3\delta_D(\mathbf{q}_{1234}),\\
&=\left(\frac{1}{N_{k_b}}\int\limits_{\vu}\Theta_{k_b}(u)a_0(u)\right)\left(\frac{1}{N_{k_b'}}\int\limits_{\vv}\Theta_{k_b'}(v)a_0(v)\right)\int\limits_{\mathbf{q}_1\dots \mathbf{q}_4}\left(\prod_{i=1}^{4}W(\mathbf{q}_i)\right)P_m^{\rm lin.}(q_{12})(2\pi)^3\delta_D(\mathbf{q}_{1234}),\\
&=\left(\frac{1}{N_{k_b}}\int\limits_{\vu}\Theta_{k_b}(u)a_0(u)\right)\left(\frac{1}{N_{k_b'}}\int\limits_{\vv}\Theta_{k_b'}(v)a_0(v)\right)\int\limits_{\mathbf{q}_{12}}|W(\vq_{12})|^2P_m^{\rm lin.}(q_{12}),\\
&=\underbrace{\left(\frac{V}{N_{k_b}}\int\limits_{\mathbf{u}}\Theta_{k_b}(u)a_0(u)\right)}_{\bar{a}_0^{k_b}}\underbrace{\left(\frac{V}{N_{k_b'}}\int\limits_{\mathbf{v}}\Theta_{k_b'}(v)a_0(v)\right)}_{\bar{a}_0^{k_b'}}\underbrace{\left[\frac{1}{V^2}\int\limits_{\mathbf{k}}|W(\mathbf{k})|^2P_m^{\rm lin.}(k)\right]}_{\sigma_W^2}.
\end{align}
where $\bar{a}_0^{k_b}$ is the leading order response of the isotropic power spectrum to a long-wavelength density perturbation,  $a_0(k)\equiv \frac{\partial P(k)}{\partial \delta_L}$, averaged over bin $k_b,$ and $\sigma_W^2$ is the variance of the linear density field over the survey volume, \emph{i.e.},
\begin{equation}\label{eq:sigma_Pk_def_window}
    \sigma_W^2=\frac{1}{V^2}\int_{\vk}|W(\vk)|^2P_m^{\rm lin}(k).
\end{equation}
To derive this result, we used $\int_{\vp}W(\vp)W(\vu-\vp)=W(\vu).$

\clearpage
\pagebreak

\section{Impact of fiducial power spectrum on the disconnected estimator}\label{App:disc_convergence}

\begin{figure*}[!t]
\centering
\includegraphics[width=0.5\linewidth]{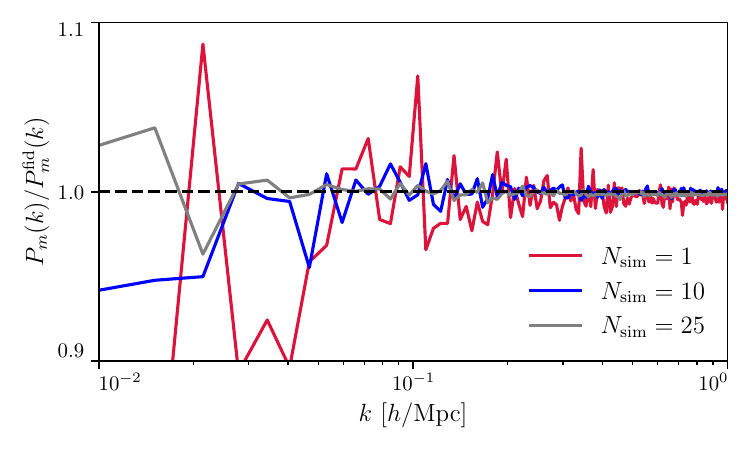}
\caption{Ratio of the mean matter power spectrum averaged over $N_{\rm sim}$ realizations to the fiducial power spectrum used in the disconnected subtraction, which uses all 15,000 \textsc{Quijote} realizations. The $\sim 5\%$ level variations in the power spectrum correspond to $\sim0.2\%$ variations in our trispectrum estimator, which is quadratic in the error in the fiducial power spectrum (see Fig.~\ref{fig:trispectrum_disc_bias}).} \label{fig:fid_Pk_disc}
\end{figure*}

To estimate the diagonal contributions to the non-Gaussian power spectrum covariance, we must subtract the disconnected contributions from the four-point function estimator, which requires a fiducial power spectrum. In this appendix, we demonstrate that our trispectrum estimator is robust to reasonable uncertainties in this fiducial spectrum.

In the main text, we assumed that the true power spectrum was known and therefore adopted the mean power spectrum measured from all 15,000 \textsc{Quijote} realizations. Here, we instead recompute the trispectrum using fiducial power spectra estimated from only 1, 10, and 25 simulations. Figure~\ref{fig:fid_Pk_disc} shows the corresponding matter power spectra. Using only a single simulation, the estimated power spectrum is accurate at roughly the $5\%$ level.

\begin{figure*}[!t]
\centering
\includegraphics[width=0.99\linewidth]{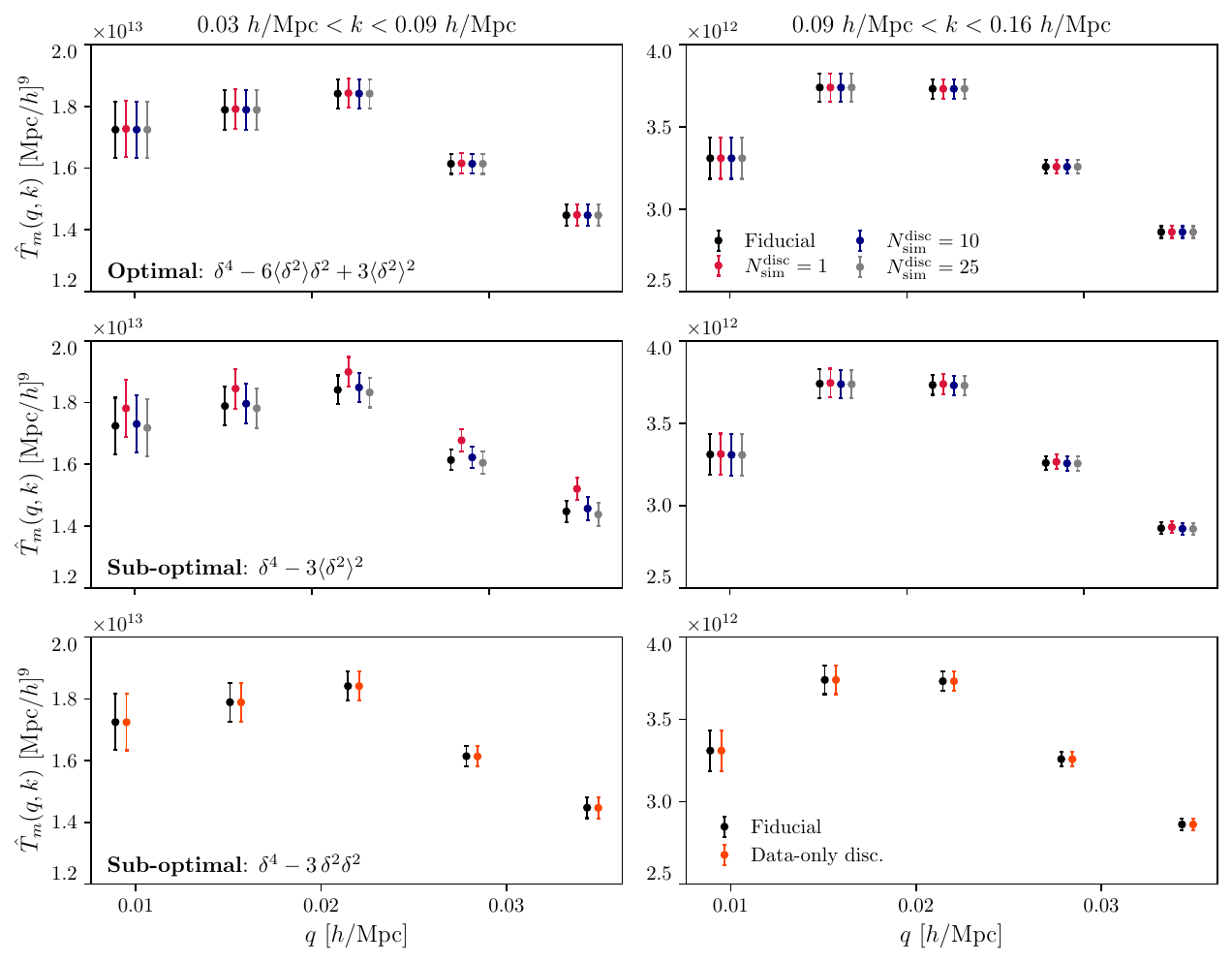}
\caption{Sensitivity of the collapsed trispectrum estimator to the disconnected four-point function estimator. We show various estimates of the collapsed trispectrum as a function of the soft mode $q$ for the two largest hard-mode bins analyzed in this work. The black points show our fiducial measurements, which use the optimal estimator with the disconnected terms estimated using the mean power spectrum from all 15,000 \textsc{Quijote} realizations. The top and middle rows show results using the optimal (realization-dependent) and sub-optimal (realization-independent) disconnected estimators, respectively. The different colors correspond to variations in the number of simulations used to estimate the disconnected term, $N_{\rm sim}^{\rm disc}$. For the optimal estimator, a single simulation is sufficient to recover the true trispectrum to within 0.2\%, because the error on the optimal estimator is quadratic in the difference between the true and fiducial power spectra. In contrast, the sub-optimal realization-independent estimator with $N_{\rm sim}^{\rm disc}=1$ is biased by $\sim 10\%$ at the largest scales. The bottom panel uses the near-optimal estimator that does not require a fiducial power spectrum (Eq.~\eqref{eq:Tdisc_data_only}). For the simulation volume and momentum bins considered here, the near-optimal estimator agrees with the optimal estimator to within $\sim 0.1\%$ across all scales. In all cases, we show the trispectrum averaged over 100 realizations.} \label{fig:trispectrum_disc_bias}
\end{figure*}

In Fig.~\ref{fig:trispectrum_disc_bias}, we compare the trispectrum estimated from 100 realizations using these three fiducial spectra with our baseline estimator. At the largest scales, using the power spectrum from a single realization, which is biased at the $\sim5\%$ level, induces only a $\sim0.2\%$ bias in the optimal trispectrum estimator, consistent with the expected quadratic scaling of the estimator bias. In contrast, if we use the sub-optimal estimator with the realization-independent disconnected term, $\delta^4 - 3\langle \delta^2\rangle^2$, we find percent-level biases in the trispectrum, as can be seen in the middle panel. Finally, using the \emph{slightly} sub-optimal estimator that does not require a fiducial power spectrum, $\delta^4 - 3\,\delta^2\,\delta^2$, we recover our fiducial trispectrum measurements to within $\sim0.1\%$ with a negligible loss in precision (bottom panels). The most significant discrepancies occur at the largest scales, where the optimal estimator is expected to out perform the sub-optimal estimator. 

In short, we do not need to use a fiducial power spectrum to estimate the disconnected terms in the present work. Nevertheless, we choose to work with the optimal estimator in the main text as we anticipate that it will prove much more useful once we account for realistic survey systematics, such as window functions.

\clearpage
\pagebreak

\section{Validation on Gaussian random fields}\label{App:GRFs}

\begin{figure*}[!t]
\centering
\includegraphics[width=0.99\linewidth]{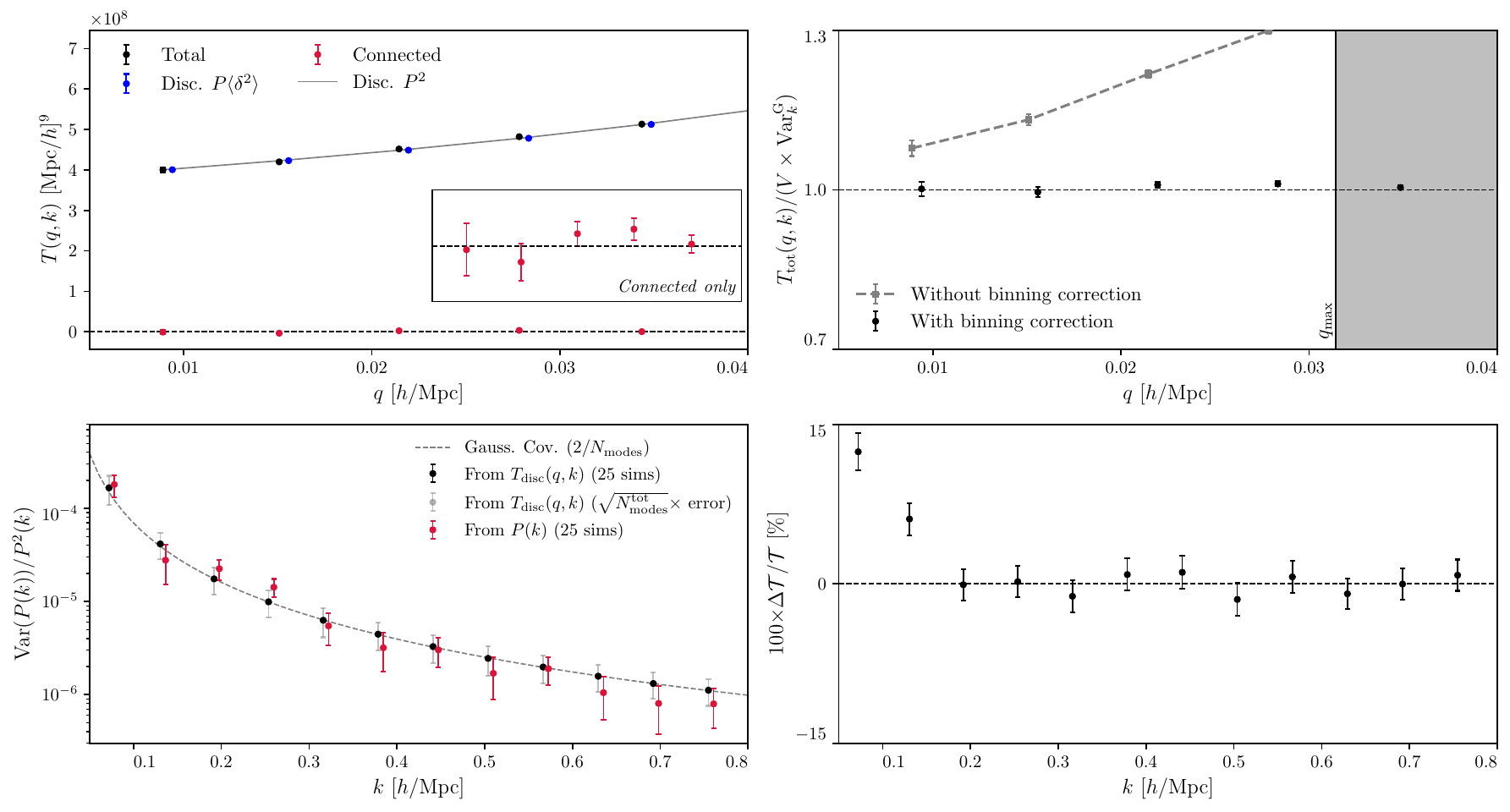}
\caption{Validation of the estimator pipeline on Gaussian random fields. The top-left panel shows the collapsed trispectrum as a function of the long-wavelength mode $q$. Black points denote the total four-point function, while the blue points and gray line show the realization-dependent and realization-independent disconnected contributions, respectively. The red points show the connected component (trispectrum), which vanishes for a Gaussian field. The top-right panel compares the total four-point function to the true Gaussian covariance. After accounting for the binning correction (Eq.~\eqref{eq:gauss_cov_GRF}), the total four-point function is completely consistent with the Gaussian covariance. Both panels are averaged over 400 simulations and correspond to the hard-mode bin centered at $k = 0.50~h/{\rm Mpc}$. The bottom-left panel shows the inferred covariance as a function of $k$, using 25 simulations. The trispectrum-based estimator yields a more precise estimate, with the improvement set directly by the number of soft modes. The bottom-right panel shows the residuals of the trispectrum-based covariance estimate to the true Gaussian covariance. The estimator recovers percent-level constraints on the Gaussian covariance for $k\gtrsim 0.15~h/{\rm Mpc}$. For $k \lesssim 0.15~h/{\rm Mpc}$, the estimator is biased due to $\mathcal{O}(q_b^2/k_b^2)$ corrections that are not included in Eq.~\eqref{eq:gauss_cov_GRF}.}
\label{fig:gauss_cov_validation}
\end{figure*}

\noindent In this appendix, we apply our four-point function covariance estimators to Gaussian random fields. As we shall see, the Gaussian covariance estimator contains many of the key ingredients used for the full non-Gaussian covariance estimation presented in the main text, and will elucidate differences between the connected and disconnected contributions to the estimators discussed in Appendix~\ref{App:trispectrum_estimators}.

Consider a 3D Gaussian random field, $\delta_{G}$, with power spectrum $P(k)$. The covariance of the binned power spectrum estimator is given by
\begin{equation}
    \cov^{\rm G}_{k_b,k_b'}=\frac{2P^2(k_b)}{N_{k_b}}\delta^{\rm K}_{k_b,k_b'}.
\end{equation}

We now consider several estimates of the power spectrum covariance. First, the sample covariance estimated from $N_{\rm sim}$ simulations using only the power spectrum follows a Wishart distribution (see, \emph{e.g.}, Ref.~\cite{Taylor:2014ota}). Consequently, the uncertainty on the sample power spectrum variance estimate is given by
\begin{equation}
      {\rm Var}[\hat{\rm Var}_{k_b}^{{\rm G};~{\rm samp}}(N_{\rm sim})]=\left(\frac{2}{N_{\rm sim}-1}\right) \left[{\rm Var}^{\rm G}_{k_b,k_b'}\right]^2=\frac{2}{N_{\rm sim}-1}\left[\frac{2 P^2(k_b)}{N_{k_b}}\right]^2,
\end{equation}
where $\hat{\rm Var}_{k_b}^{G;~{\rm samp}}(N_{\rm sim})$ denotes the estimated sample variance from $N_{\rm sims}$. 

We now consider trispectrum-based covariance estimation. Since the trispectrum of a Gaussian random field vanishes, we work directly with the total four-point function estimator in Eq.~\eqref{eq:total_estimator_Tk}, evaluated for a non-zero internal mode, $Q_b \neq 0$. Our goal is to relate the expectation value of this estimator to the Gaussian power spectrum covariance. Because the Gaussian covariance is diagonal, we take $k_{b_1} = k_{b_2} = k_{b_3} = k_{b_4} = k_b$, yielding
\begin{align}
    \hat{T}_{\rm tot.}(k_b, Q_b)\propto \int\limits_{\vP}\Theta_{Q_b}(P)\int\limits_{\vp_1,\dots, \vp_4}\bigg[&\prod_{i=1}^4\Theta_{k_b}(p_i)\bigg]\, \delta_{\vp_1}\delta_{\vp_2}\delta_{\vp_3}\delta_{\vp_4}(2\pi)^3\delta_D^{(3)}(\vp_{12}-\vP)(2\pi)^3\delta_D^{(3)}(\vp_{34}+\vP)\nonumber.
\end{align}
Taking the expectation value for a Gaussian random field, $\langle \delta_{\vp_1}\delta_{\vp_2}\delta_{\vp_3}\delta_{\vp_4}\rangle=(2\pi)^6\delta_D^{(3)}(\vp_{12})\delta_D^{(3)}(\vp_{34})P(p_1)P(p_3)+{\rm 2~perm.}$, and noting that the first permutation does not contribute when $Q_b\neq 0$, we find
\begin{align}\label{eq:tot_4pcf_gauss_expectation}
    \hat{T}_{\rm tot.}(k_b,Q_b)&\propto 2(2\pi)^{12}\int\limits_{\vP}\Theta_{Q_b}(P)\int\limits_{\vp_1,\dots, \vp_4}\bigg[\prod_{i=1}^4\Theta_{k_b}(p_i)\bigg]\delta_D^{(3)}(\vp_{13})\delta_D^{(3)}(\vp_{24})P(p_1)P(p_2)\delta_D^{(3)}(\vp_{12}-\vP)\delta_D^{(3)}(\vp_{34}+\vP)\nonumber,\\
    &\propto 2\int\limits_{\vP}\Theta_{Q_b}(P)\int\limits_{\vp_1,\vp_2}\Theta_{k_b}(p_1)\Theta_{k_b}(p_2)P(p_1)P(p_2)\left[(2\pi)^3\delta_D^{(3)}(\vp_{12}-\vP) \right]^2,\nonumber\\
     &\propto 2V\int\limits_{\vp_1,\vp_2}\Theta_{Q_b}(p_{12})\Theta_{k_b}(p_1)\Theta_{k_b}(p_2)P(p_1)P(p_2).
\end{align}
Eq.~\eqref{eq:tot_4pcf_gauss_expectation} closely resembles the squared, bin-averaged power spectrum that appears in the Gaussian covariance. The key difference is the presence of a $q$-dependent weighting, given by the triangle-counting factor used to normalize the bispectrum estimator. This weighting induces a $q$-dependence in the total four-point function estimator, such that the power spectrum covariance of a Gaussian random field can be written in terms of the total four-point function of the field as follows,
\begin{equation}\label{eq:gauss_cov_GRF}
    {\rm Cov}_{\rm G}(k_b,k_b') = \frac{1}{V} \left[ \hat{T}_{\rm tot.}(q_b,k_b)\,\frac{N_{\rm tri}(q_b,k_b)}{N_{q_b}N_{k_b}} \right] \delta^{\rm K}_{k,k'}+\mathcal{O}(q_b^2/k_b^2),
\end{equation}
where the $\mathcal{O}(q_b^2/k_b^2)$ are sub-dominant corrections arising from variations in the power spectrum within a bin. 

To test Eq.~\eqref{eq:gauss_cov_GRF}, we generate 400 Gaussian random fields using the linear matter power spectrum at $z=0$ for the fiducial \textsc{Quijote} cosmology with a $(1~{\rm Gpc}/h)^3$ volume and $N_{\rm grid}^3=512^3$. We then compute the total four-point function of the Gaussian random fields using the same bins as in the main text, \emph{i.e.}, five soft-mode bins between $0.006~h/{\rm Mpc}\simeq k_f\leq q_b< 6k_f\simeq 0.037~h/{\rm Mpc}$ with bin spacing equal to the fundamental mode, and twelve hard-mode bins between $5k_f \simeq 0.03~h/{\rm Mpc}\lesssim k_b\lesssim 120k_f\simeq 0.75~h/{\rm Mpc}$ with width $10k_f\simeq0.06~h/{\rm Mpc}$. 

The top-left panel of Fig.~\ref{fig:gauss_cov_validation} show various contributions to the trispectrum estimator as a function of the soft mode for the hard-mode bin centered at $k\sim 0.50~h/{\rm Mpc}$, averaged over all four hundred Gaussian simulations. The black points show the total four-point function estimate. The blue points and gray lines show the realization-dependent and realization-independent disconnected estimators, respectively. The red points show the connected four-point function, \emph{i.e.}, the trispectrum, computed using the optimal estimator described in Appendix~\ref{App:trispectrum_estimators}. The trispectrum of the Gaussian random fields is completely consistent with zero, as can be seen in the inset panel. 

In the top-right panel of Fig.~\ref{fig:gauss_cov_validation}, we plot the ratio of the total four-point function in the left panel to the theoretical prediction for the Gaussian power spectrum covariance. After accounting for the binning correction (multiplying $\hat{T}_{\rm tot}$ by $N_{\rm tri}(q_b,k_b)/N_{q_b}N_{k_b}$), the total four-point function exactly matches the Gaussian covariance. 

Having established the correspondence between the total four-point function estimator and the power spectrum covariance, for Gaussian random fields, we fit for the Gaussian covariance using the pipeline developed in the main text. In the bottom-left panel of Fig.~\ref{fig:gauss_cov_validation}, we show the estimated power spectrum covariance as a function of the hard-mode bin using 25 simulations. The trispectrum-based estimator (black points) recovers the true covariance to within a few percent. Averaging over all $k$ bins, we find that the trispectrum-based estimator outperforms the sample covariance by a factor of $21.4\pm 1.0$, which is completely consistent with the mode counting expectation of $\sqrt{N_{\rm modes}^{\rm tot}}=22$. 

Finally, the bottom-right panel of Fig.~\ref{fig:gauss_cov_validation} shows the residuals between the estimated and true covariance as a function of the hard-mode bin. For $k \gtrsim 0.15~h/{\rm Mpc}$, the estimator is unbiased. At lower $k$, the estimated covariance is biased due to $\mathcal{O}(q_b^2/k_b^2)$ corrections that are not included in Eq.~\eqref{eq:gauss_cov_GRF}.

\clearpage
\pagebreak

\section{Additional details regarding the likelihood and fitting procedure}\label{App:likelihood}
%
\begin{figure}[!t]
\centering
\subfloat[$N_{\rm sim}^{\rm cov}=25$]{%
    \includegraphics[width=0.48\linewidth]{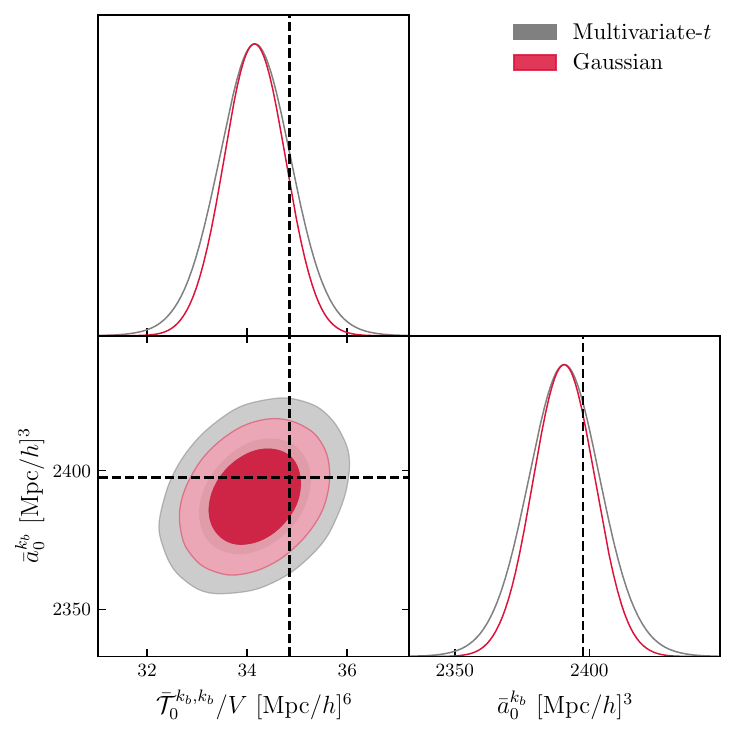}
    \label{fig:vary_likelihood_a}
}
\hfill
\subfloat[$N_{\rm sim}^{\rm cov}=200$]{%
    \includegraphics[width=0.48\linewidth]{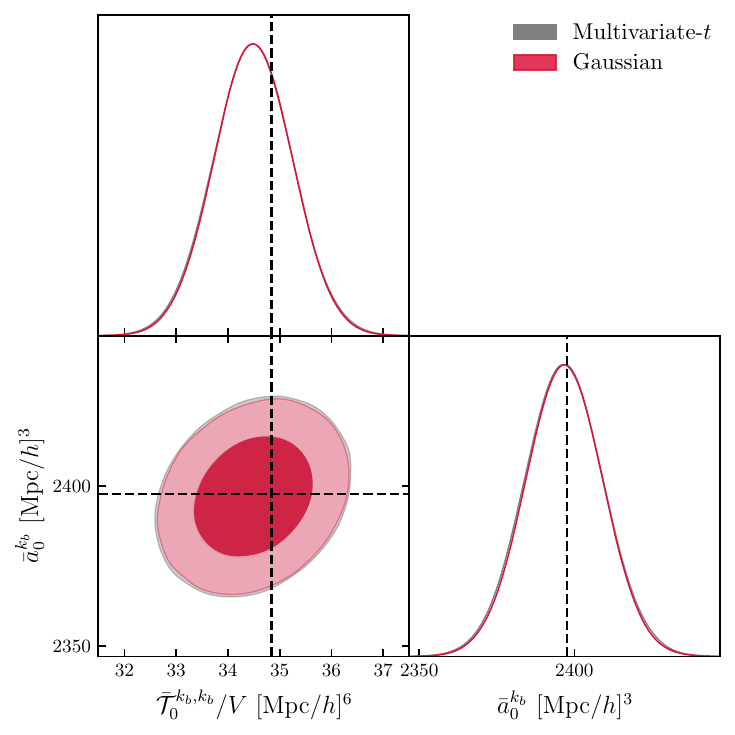}
    \label{fig:vary_likelihood_b}
}

\caption{Impact of the choice of likelihood on the inferred parameters governing the power spectrum covariance. In our fiducial analysis, we use a multivariate-$t$ likelihood to account for the finite number of simulations $(N_{\rm sim}^{\rm cov}=25)$ used to estimate the power spectrum, squeezed bispectrum, and collapsed trispectrum covariance. If we assume a Gaussian likelihood, we underestimate the uncertainty on $\bar{a}_0^{k}$ and $\bar{\mathcal{T}}^0_{k,k'}$ by 20\%. The two likelihoods give indistinguishable constraints if we use more simulations to estimate the covariance, as illustrated in panel (b), which uses $N_{\rm sim}^{\rm cov}=200$.  These plots use measurements centered at $k_b=0.50~h/{\rm Mpc}$.}
\label{fig:vary_likelihood}
\end{figure}

In our fiducial analysis, we use a multivariate-$t$ likelihood (Eq.~\eqref{eq:likelihood}) to account for the fact that the power spectrum, bispectrum, and trispectrum covariance matrices are estimated from a finite number of simulations ($N_{\rm sim}^{\rm cov}=25$)~\cite{Sellentin:2015waz}. In Fig.~\ref{fig:vary_likelihood}, we compare our constraints using a multivariate-$t$ likelihood to those obtained assuming a Gaussian likelihood. The left (right) panel shows results using 25 (200) simulations to estimate the covariance. For $N_{\rm sim}^{\rm cov}=25$, the Gaussian likelihood underestimates parameter uncertainties by $\sim 20\%$ relative to the multivariate-$t$ distribution. In contrast, for $N_{\rm sim}^{\rm cov}=200$, the two likelihoods yield indistinguishable results, as expected since Eq.~\eqref{eq:likelihood} reduces to a Gaussian likelihood in the $N_{\rm sim}\to\infty$ limit. 

\begin{figure*}[!t]
\centering
\includegraphics[width=0.7\linewidth]{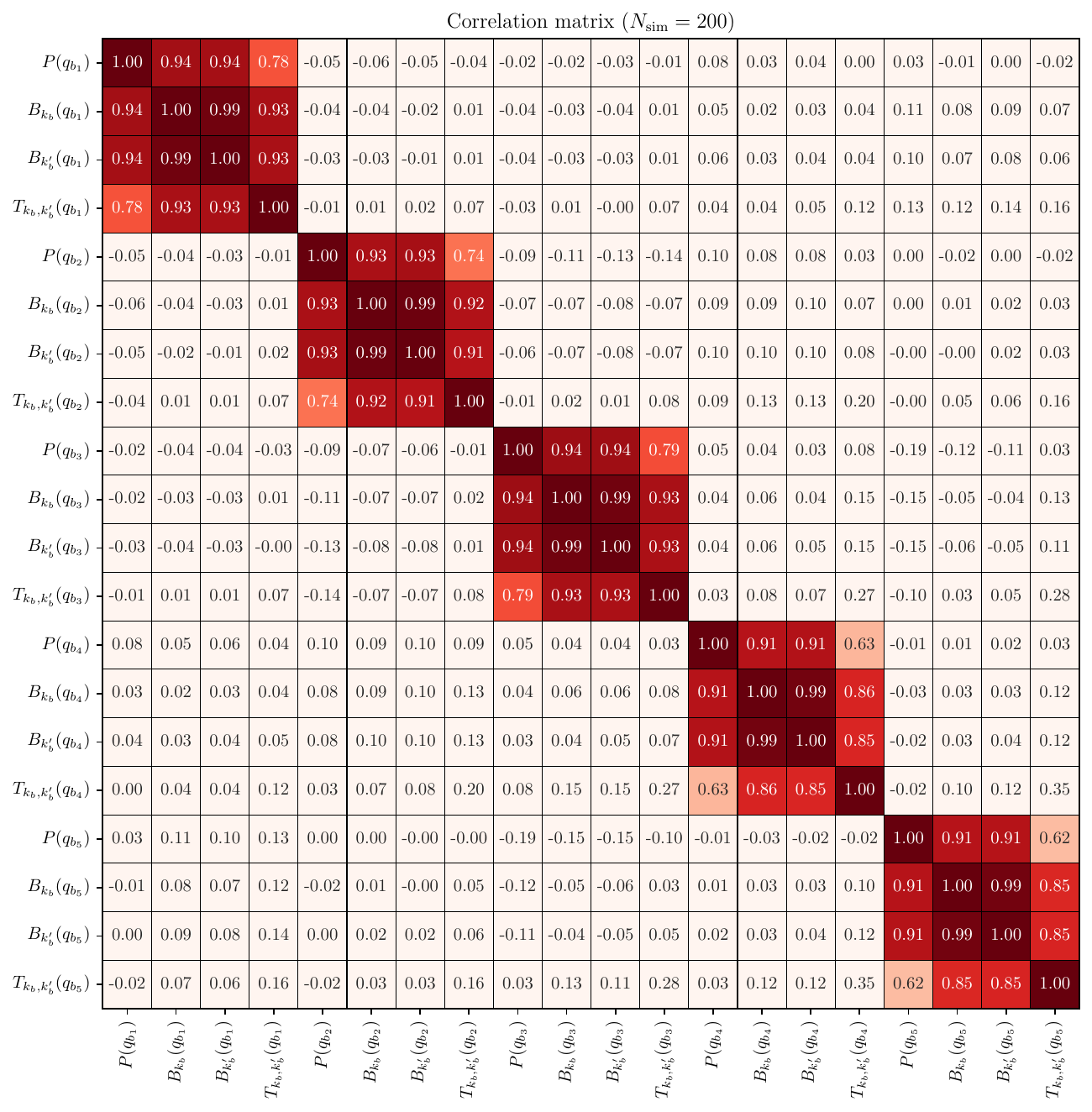}
\caption{Correlation matrix of the power spectrum, squeezed bispectrum, and collapsed trispectrum, grouped by soft-mode bin $q_{b_i}$. For each $q_{b_i}$, we show the long-wavelength power spectrum, the squeezed bispectrum evaluated in two distinct hard-mode bins ($k_b \neq k_b'$), and the collapsed trispectrum in the asymmetric parallelogram configuration, with $k_b \approx 0.38~h/{\rm Mpc}$ and $k_b' \approx 0.50~h/{\rm Mpc}$. Within a fixed $q_{b_i}$ bin the four statistics are highly correlated, as they probe the same underlying long-wavelength modes. Conversely, the correlations between different $q$ bins are negligible. Therefore, in the main text, we approximate the covariance as block-diagonal.} \label{fig:full_corr_mat}
\end{figure*}

In our fiducial analysis, we also assume that long-wavelength power spectrum, squeezed bispectrum, and collapsed trispectrum are uncorrelated for different long-wavelength mode bins, $q_{b_i}.$ To assess the validity of this assumption, Fig.~\ref{fig:full_corr_mat} shows the full correlation matrix of the power spectrum, squeezed bispectrum, and collapsed trispectrum in different soft-mode bins $q_{b_i}$, with fixed hard modes bins centered at $k_{b}=0.38~h/{\rm Mpc}$ and $k_{b'}=0.50~h/{\rm Mpc}$. Whereas the power spectrum, bispectrum, and trispectrum are highly correlated within the same soft-mode bin, since these statistics trace the same realization of the underlying long-wavelength modes (see Fig.~\ref{fig:PBT_binned}), they are largely uncorrelated between different soft-mode bins. The strongest correlations between soft-mode bins are in the collapsed trispectrum, with the $q_{b_4}\approx 0.028~h/{\rm Mpc}$ and $q_{b_5}\approx 0.034~h/{\rm Mpc}$ bins correlated at over 30\%. We note that in the main text we fix $q_{\rm max}=0.03~h/{\rm Mpc}$, hence we do not include the $q_{b_5}$ bin in our analysis. 

\begin{figure*}[!t]
\centering
\includegraphics[width=0.5\linewidth]{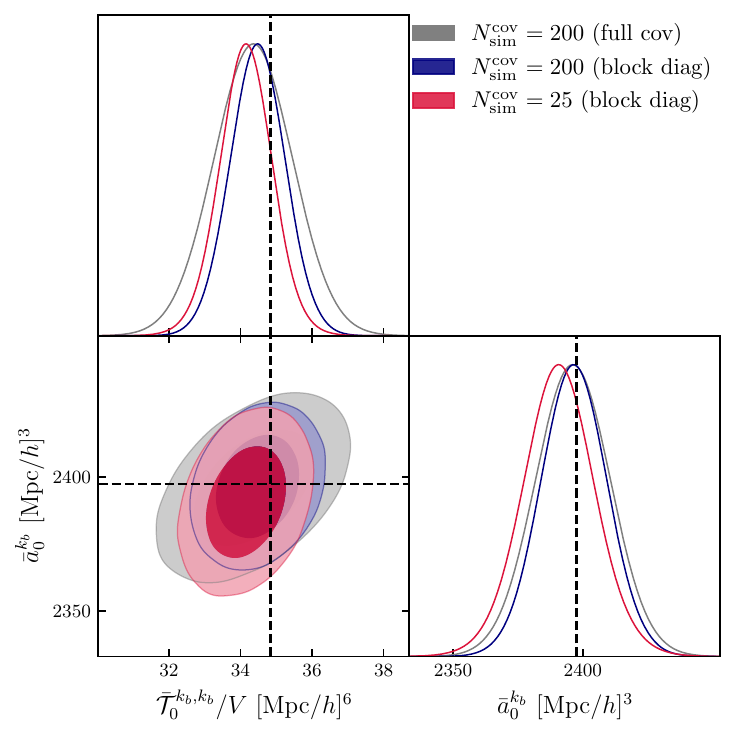}
\caption{Impact of the covariance estimate for the power spectrum, bispectrum, and trispectrum on constraints of the non-Gaussian power spectrum covariance parameters. In our fiducial analysis (red), we adopt a block-diagonal covariance estimated from 25 simulations. These constraints are consistent within $<0.4\sigma$ of those obtained using a block-diagonal covariance estimated from 200 simulations. Relaxing the block-diagonal assumption increases the uncertainty on $\mathcal{T}^0_{k,k'}$ by $\sim 30\%$, demonstrating that correlations of the trispectrum across soft-mode bins have a relatively small impact on the final constraints. This plot uses measurements centered at $k_b=0.50~h/{\rm Mpc}$.} \label{fig:vary_fid_cov}
\end{figure*}

Based on Fig.~\ref{fig:full_corr_mat}, the block-diagonal covariance provides a reasonable approximation for our main analysis. To further validate this assumption, we repeat the analysis using the full covariance estimated from 200 simulations. In Fig.~\ref{fig:vary_fid_cov}, we compare constraints obtained with the full covariance to those using a block-diagonal approximation, focusing on the small-scale bin centered at $k_b=0.50~h/{\rm Mpc}$. The red and blue contours correspond to block-diagonal covariances estimated from 25 and 200 simulations, respectively, and are consistent within $\lesssim 0.4\sigma$, indicating that the block-diagonal estimate is sufficiently converged with only 25 simulations. The gray contours show results obtained using the full covariance estimated from 200 simulations. While the constraint on $\bar{a}_0^{k}$, which is dominated by the squeezed bispectrum, remains unchanged, the uncertainty on $\bar{\mathcal{T}}^0_{k,k}$ increases by $\sim 30\%$, due to the mild correlations in the trispectrum across soft-mode bins. In this work, we neglect these correlations, as they are subdominant compared to the overall improvement of our power spectrum covariance estimator compared to the sample covariance. Nevertheless, in the future, we hope to generalize our approach such that it does not require simulations to estimate the trispectrum covariance. 

\clearpage
\pagebreak

\section{Full covariance matrix}\label{App:full_covariance_matrix}

\begin{figure*}[!t]
\centering
\includegraphics[width=0.99\linewidth]{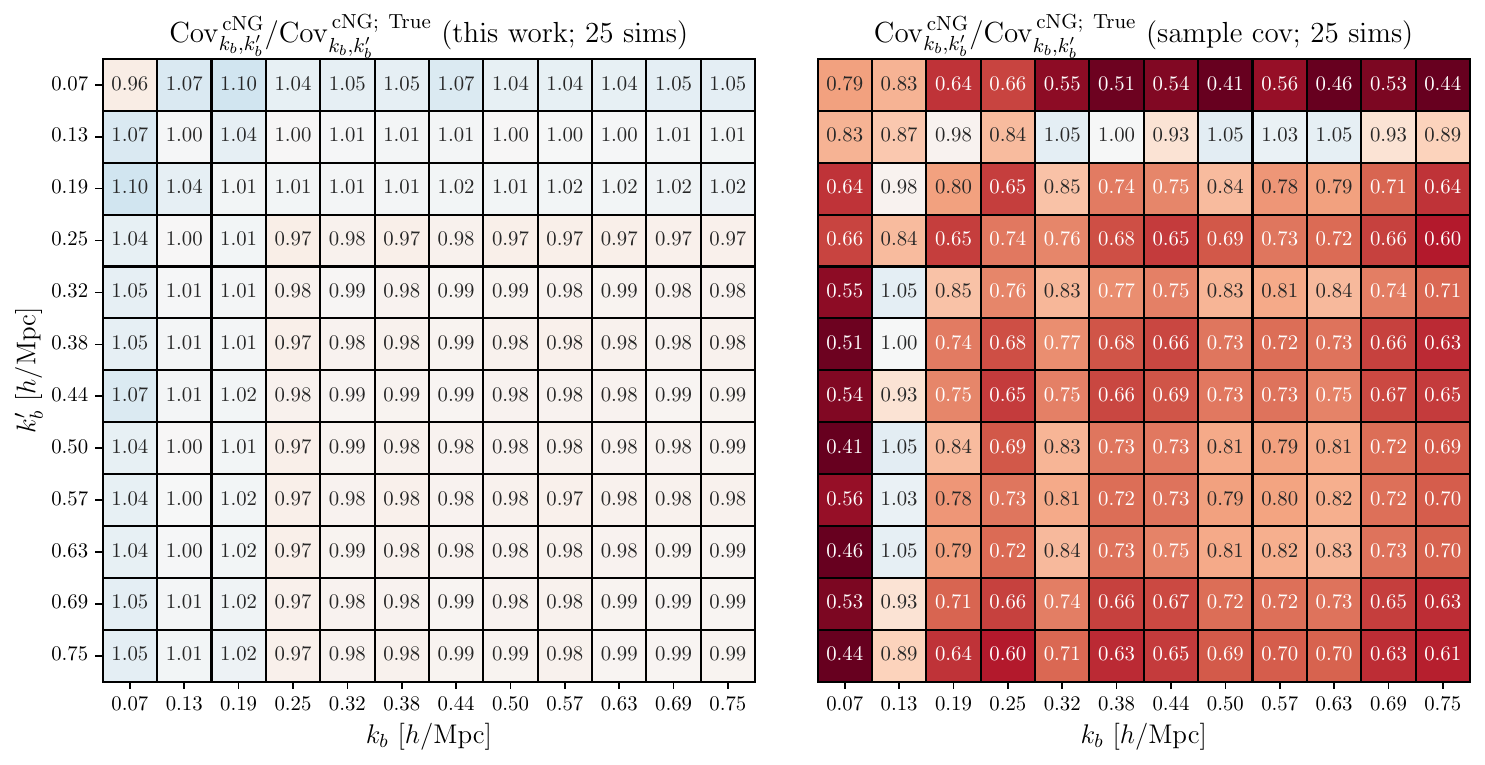}
\caption{Accuracy of the connected non-Gaussian covariance estimated from the squeezed bispectrum and collapsed trispectrum (left) and from the standard sample covariance (right), using 25 simulations. In each panel, we show the ratio relative to the “true” connected non-Gaussian covariance, estimated from 15,000 simulations. Our method recovers the covariance to within a few percent across all scales analyzed, except in the lowest $k$ bin, where the trispectrum is not sufficiently collapsed.} \label{fig:full_cov_ratio}
\end{figure*}

 Fig.~\ref{fig:full_cov_ratio} shows the accuracy of the connected non-Gaussian covariance estimated from the squeezed bispectrum and collapsed trispectrum (left) and the usual sample covariance (right), with 25 simulations.

\end{document}